\documentclass[letterpaper,twocolumn,10pt]{article}
\usepackage{USENIX_Style}

\usepackage{amsmath}
\usepackage{todonotes}
\usepackage{cleveref}
\usepackage{subcaption}
\usepackage{bbm}
\usepackage{booktabs} 
\usepackage{algorithm}
\usepackage{algpseudocode}
\usepackage{url}
\usepackage{xcolor}
\usepackage[normalem]{ulem}

\DeclareGraphicsExtensions{.pdf}
\graphicspath{{images/pdf/}{images/png/}}

\usepackage{filecontents}

\usepackage{bibspacing}

\newcommand{\Description}[1]

\usepackage{alphalph}
\usepackage{titlesec}

\titlespacing{\section}{0pt}{12pt plus 4pt minus 2pt}{4pt}
\titlespacing{\subsection}{0pt}{6pt plus 2pt minus 2pt}{0pt}
\titlespacing{\subsubsection}{0pt}{6pt plus 4pt minus 2pt}{0pt}
\titlespacing{\paragraph}{0pt}{2pt plus 2pt minus 0pt}{10pt}

\begin{document}
\date{}

\title{\Large \bf 
Security Analysis of Camera-LiDAR Fusion Against Black-Box Attacks on Autonomous Vehicles}


\author{
{\rm R. Spencer Hallyburton}\\
Duke University
\and
{\rm Yupei Liu}\\
Duke University
\and
{\rm Yulong Cao}\\
University of Michigan
\and
{\rm Z. Morley Mao}\\
University of Michigan
\and
{\rm Miroslav Pajic}\\
Duke University
}

\maketitle

\begin{abstract}
To enable safe and reliable decision-making, autonomous vehicles (AVs) feed sensor data to \textit{perception} algorithms to understand the environment. Sensor fusion with multi-frame tracking is becoming increasingly popular for detecting 3D objects. Thus, in this work, we perform an analysis of camera-LiDAR fusion, in the AV context, under LiDAR spoofing attacks. Recently, LiDAR-only perception was shown vulnerable to LiDAR spoofing attacks; however, we demonstrate these attacks are not capable of disrupting camera-LiDAR fusion. We then define a novel, context-aware attack: \emph{frustum attack}, and show that out of 8 widely used perception algorithms -- across 3 architectures of LiDAR-only and 3 architectures of camera-LiDAR fusion -- \emph{all} are significantly~vulnerable to the frustum attack. In addition, we demonstrate that the frustum attack is stealthy to existing defenses against LiDAR spoofing as it preserves consistencies between camera and LiDAR semantics. Finally, we show that the frustum attack can be exercised consistently over time to form stealthy longitudinal attack sequences, compromising the tracking module and creating adverse outcomes on end-to-end AV control.
\end{abstract}

\section{Introduction}
\label{sec:1-intro}
\vspace{-2pt}

Autonomous vehicles (AVs) have enjoyed millions of miles of partially automated road travel~\cite{Hawkins2018, 2021TheTechnologies}. 
This has been enabled by advances in \emph{perception}, the foundation for safe and reliable decision-making in AVs. Sensors including cameras and light detection and ranging (LiDAR) collect data so perception can provide AVs enough awareness of surroundings to make informed decisions in safety-critical tasks such as obstacle/pedestrian avoidance and traffic sign~detection.

The camera and LiDAR are the most used AV sensors~\cite{Hecht2018, 2017GMDeveloper, 2021NVIDIADRIVE, 2021BaiduApollo}. Inexpensive, high-quality cameras can provide high resolution, dense 2D outputs on limited fields of view. LiDAR is complementary to the camera, providing up to $360^{\circ}$ view of the surroundings and fully resolving the 3D position of objects with a sparse set of points (i.e., point~clouds).

Due to AVs' safety-critical nature, misinformation or wrong decisions can quickly lead to severe adverse outcomes~\cite{Schoettle2015, Kohli2019}. The high-impact outcomes underscore the need for security research in the domain. In particular, the increasing reliance of AVs on deep neural networks (DNNs) for real-time perception has sparked security questions at the algorithm level. There is a growing body of AV perception security work: an attacker can perturb sensor data to change object classification (misclassification)~\cite{Eykholt2018b}, introduce fake objects (false positives)~\cite{Cao2019d, Sun2020h}, and remove existing objects (false negative)~\cite{Tu2021b, Abdelfattah2021}, each with devastating consequences at the driving decision and control levels.

Initially, security analysis of perception focused on the image domain with LiDAR only recently emerging as the target for security research. Spoofing attacks against LiDAR have since been demonstrated~\cite{Petit2015c, Shin2017d, Cao2019d, Sun2020h, Cao2021AutomatedTargets}, and applied to LiDAR-only perception~\cite{Cao2019d, Sun2020h}. The use of physical adversarial objects has also been explored~\cite{Tu2021b, Cao2021InvisibleAttacks, Abdelfattah2021}, demonstrating outcomes against end-to-end AV pipelines~\cite{Cao2021InvisibleAttacks}.

However, existing security analyses of LiDAR-based perception have several limitations. Reported physically-realizable attacks mainly consider single-sensor (e.g.,~LiDAR-only, camera-only) perception. On the other hand, deployed AV architectures such as Waymo's One~\cite{Hawkins2018}, 
Baidu's Apollo~\cite{2021BaiduApollo}, and NVIDIA's DRIVE~\cite{2021NVIDIADRIVE} employ multi-sensor perception with multi-frame tracking. 
Security analysis of multi-sensor fusion has been recently considered~\cite{Tu2021b, Cao2021InvisibleAttacks, Abdelfattah2021}; e.g.,~\cite{Cao2021InvisibleAttacks} focuses on the impact of adversarial physical objects on camera-LiDAR  perception. Yet, these approaches require highly representative models of the deployed perception algorithms to design attacks with white-box optimization online or a-priori. To the best of our knowledge, there is no analysis of black-box (i.e., when the perception model is not known to the attackers) attacks against sensor-fusion perception.

Consequently, in this work, \emph{we present security analysis for camera-LiDAR sensor fusion under physically-demonstrated black-box LiDAR spoofing attacks}. Using the LiDAR-spoofing threat model from~\cite{Cao2019d, Sun2020h}, we first show that \emph{camera-LiDAR fusion confers additional robustness against general black-box (i.e.,~naive) LiDAR attacks}; this is because the naive spoofing does not retain consistency between camera data and thus can be filtered. That attack success may be greatly reduced with sensor fusion when not all sensors are compromised has been suggested in prior works~\cite{Liu2021, Cao2019d, Sun2020h, Tu2021b, Pajic2014b}, and is systematically evaluated for the first time in this work.

Unlike the recent work of~\cite{Liu2021} that restricts analysis to naive LiDAR attacks, we introduce a new class of perception attacks, the \emph{frustum attack}, which compromises camera-LiDAR fusion by preserving semantic consistencies between the camera and LiDAR data. To achieve this, the attacker only needs to know approximate locations of true objects in the scene. We experimentally demonstrate that the frustum attack can be executed in the physical world with limited knowledge. We describe five scenarios where an adversary can use contextual information to \emph{launch spoofing attacks relative to existing objects in the scene}. This expands upon prior works~\cite{Cao2019d, Sun2020h} that focused only on naive, isolated placement at 5-8~m~range.

We then evaluate the frustum attack against state-of-the art defenses against LiDAR spoofing~\cite{Sun2020h, Hau2020Shadow-Catcher:Sensing, Liu2021} using a diverse set of eight perception algorithms across three distinct LiDAR-only and three distinct camera-LiDAR fusion architectures, including cascaded-semantic, feature-level, and tracking-level fusions (Fig.~\ref{fig:fusion-archs}) on over 75 million attack scenarios. To the best of our knowledge, this constitutes the largest analysis of LiDAR spoofing to date and the first that extensively evaluates multiple architectures of multi-sensor fusion for perception.

In addition to false positives (FPs), we demonstrate that the frustum attack is successful in generating false negative (FN) and translation attack outcomes, as defined in Section~\ref{sec:3-attack-model-attack-goals}, which is a novel discovery for LiDAR spoofing attacks.

We also show that a key assumption about the required attacker's capabilities from prior work can be relaxed. Existing spoofing attacks have only had success at creating FPs or FNs with precise (cm-level) placement of points; furthermore, existing LiDAR spoofing attacks have required either white-box model access~\cite{Cao2019d} or carefully-crafted point placements in the outline of real vehicles (e.g.,~adversary pre-captures samples and replays them~\cite{Sun2020h}). We establish that \emph{inserting a random sample of normally-distributed points is comparably as successful as inserting points in the outline of a car.} This confers inherent attack robustness to small perturbations, facilitating attack deployment with a physical spoofing device such as in~(e.g.,~\cite{Petit2015c, Shin2017d, Cao2019d, Cao2021AutomatedTargets}), and as demonstrated in Section~\ref{sec:6-frustum-attack-experiment-details}. 

Finally, to assess the impact of LiDAR attacks on AVs equipped with multi-frame tracking, we present frustum attack case studies using longitudinal sequences of perception data. First, we explicitly analyze the multi-frame fusion and tracking module, which is employed by all industry AVs, using representative algorithms. Then, we test the frustum attack on Baidu Apollo~\cite{2021BaiduApollo} using the LGSVL simulator~\cite{Rong2020LgsvlDriving}. The case studies illuminate that high-impact adversarial situations that endanger vehicle and passenger safety occur under the frustum attack when attacking over multiple time~points, effectively deceiving the host vehicle's tracking and control.

In summary, we make the following main contributions: 

\noindent$\bullet$ We show that several sensor-fusion algorithms are robust to naive LiDAR spoofing at some of the highest defensive rates yet observed (e.g.,  $<1\%$ for some algorithms), suggesting sensor fusion is inherently secure against naive~attacks.

\noindent$\bullet$ We introduce a novel class of LiDAR spoofing attacks on AVs, the \emph{frustum attack}, and experimentally validate frustum attack feasibility with existing hardware.

\noindent$\bullet$ We perform a thorough analysis of LiDAR-only and camera-LiDAR perception and show the frustum attack's first-of-a-kind ability to compromise 8 high-performing perception algorithms across 3 LiDAR-only and 3 camera-LiDAR fusion architectures. We also show that the \emph{{frustum attack is stealthy even against existing defenses of LiDAR~spoofing}}.

\noindent $\bullet$ We perform longitudinal studies of security against perception attacks. We show that, on an end-to-end AV driving software, by using frustum attacks to fool the AV's tracking and control, the attacker has high levels of attack success attacking at short and long range, expanding on previous short range attack cases.
\vspace{-6pt}
\section{Background and Related Work}
\label{sec:2-background}
\vspace{-2pt}

\subsection{Perception} \label{sec:perception}
AVs interact in complex environments with active agents and dynamic weather and terrain situations. To accomplish desired tasks while retaining consistent situational awareness, deployed AVs are equipped with multiple sensors of multiple modalities as well as with perception algorithms to translate sensor data into meaningful semantic information  (e.g.,~vehicle tracking for situational awareness). 

\subsubsection{Camera and LiDAR Sensing}
AVs are equipped with multiple cameras spaced around the vehicle. Individual cameras provide monocular~vision which resolve azimuth and elevation angles to targets. Cameras are inexpensive compared to LiDAR and radar and thus are the preferred sensing modality for many AVs \cite{2021NVIDIADRIVE, Geiger2013b, 2021BaiduApollo, Hecht2018, Kato2018AutowareSystems}. 

A central scanning LiDAR is commonly mounted on the roof of 
AVs for maximum viewing opportunity. LiDAR is complementary to the camera; it is an active sensor that sends primarily infrared light and constructs transmit-receive time differences to resolve the full 3D position of point returns~\cite{Hecht2018}. LiDAR has demonstrated enhanced robustness compared to cameras in situations including adverse weather~\cite{Kutila2016}.

\subsubsection{AV Benchmarks}
We use KITTI~\cite{Geiger2013b} and the LGSVL Simulator~\cite{Rong2020LgsvlDriving} to test our algorithms and attacks. KITTI is composed of synchronized camera and LiDAR captures with ground truth 2D and 3D bounding boxes. We use perception algorithms with publicly available models pretrained on KITTI (see Sec.~\ref{sec:percepionAlgs}) as well as Baidu Apollo's open source end-to-end AV stack~\cite{2021BaiduApollo}.

\subsubsection{Perception Algorithms}
\label{sec:percepionAlgs}

\begin{figure*}[!t]
\begin{subfigure}[t]{.3\linewidth}
  \centering
  \includegraphics[width=0.7\linewidth]{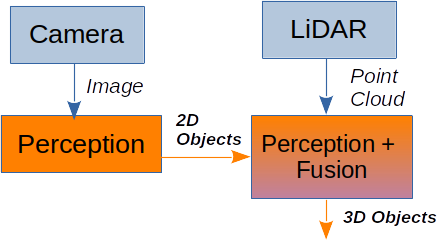}
  \vspace{-8pt}
  \caption{Cascaded semantic-level fusion}
  \label{fig:fusion-cascaded}
  \vspace{-12pt}
\end{subfigure}
\begin{subfigure}[t]{.42\linewidth}
  \centering
  \includegraphics[width=0.48\linewidth]{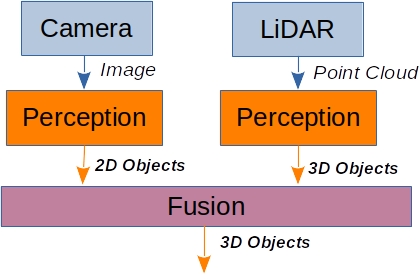}
  \vspace{-8pt}
  \caption{Integrated semantic-level fusion (e.g.,~fusion at tracking)}
  \label{fig:fusion-integrated}
  \vspace{-12pt}
\end{subfigure}
\begin{subfigure}[t]{.3\linewidth}
  \centering
  \includegraphics[width=0.7\linewidth]{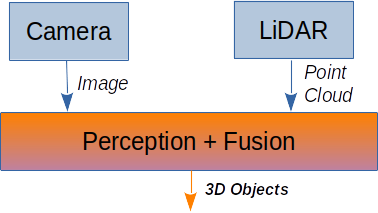}
  \vspace{-8pt}
  \caption{Feature-level fusion}
  \label{fig:fusion-feature}
  \vspace{-12pt}
\end{subfigure}
\vspace{-12pt}
\caption{The three widely-used classes of multi-sensor fusion in perception.}
\vspace{-10pt}
\label{fig:fusion-archs}
\end{figure*}

Recently, novel DNN-based methods have been proposed for processing point cloud data from LiDAR. Three general classes of LiDAR-only perception were reported in~\cite{Sun2020h} and include the bird's-eye view (BEV) (e.g.,~\cite{Liang2018a, 2021BaiduApollo}), voxelization of the 3D space (e.g.,~\cite{Zhou2018a, Lang2019a}), and direct ingesting of points (e.g.,~\cite{Shi2019}). These early works focused on single-sensor perception pipelines without considering multi-sensor~fusion.

We consider three broad classes of multi-sensor fusion for perception, illustrated in Fig.~\ref{fig:fusion-archs}: (1)~cascaded semantic fusion (e.g.,~\cite{Qi2018b, Wang2019c}) using the output of perception on one or more sensors to augment the input of other single-sensor perception, (2)~integrated semantic fusion (e.g.,~\cite{2021BaiduApollo}) that runs isolated perception for each sensor and fuses semantic outputs (e.g.,~in tracking), and (3)~feature-level fusion (e.g.,~\cite{Ku2018, Huang2020Epnet:Detection}) that combines low-level (machine-learned) features from multiple perception sources to produce a unified output. Specifically, we analyze state-of-the-art LiDAR-only (PointPillars~\cite{Lang2019a}, PointRCNN~\cite{Shi2019}, PIXOR~\cite{Yang2018a}) and camera-LiDAR fusion (Frustum-ConvNet~\cite{Wang2019c}, Frustum-PoinNet~\cite{Qi2018b}, AVOD~\cite{Ku2018}, EPNET~\cite{Huang2020Epnet:Detection}, Baidu Apollo~\cite{2021BaiduApollo}) perception algorithms, as summarized in Table~\ref{tab:perception-algs}. 

\begin{table}[!t]
  \caption{LiDAR-only and camera-LiDAR fusion perception algorithms of multiple architectures are all evaluated under the naive and frustum attacks and with defenses.}
\vspace{-8pt}
  \centering
  \label{tab:perception-algs}
  \begin{tabular}[t]{lcc}
    \toprule
    \textbf{Algorithm} & \textbf{Type} & \textbf{Architecture}\\
    \midrule
    PointPillars~\cite{Lang2019a} & LiDAR & Voxel-Based\\
    PointRCNN~\cite{Shi2019} & LiDAR & Point-Based\\
    PIXOR~\cite{Yang2018a} & LiDAR & BEV\\
    \midrule
    Frust.-ConvNet~\cite{Wang2019c} & Camera-LiDAR & Casc. Semantic\\
    Frust.-PointNet~\cite{Qi2018b} & Camera-LiDAR & Casc. Semantic\\
    AVOD~\cite{Ku2018} & Camera-LiDAR & Feature-Level\\
    EPNET~\cite{Huang2020Epnet:Detection} & Camera-LiDAR & Feature-Level\\
    Baidu Apollo~\cite{2021BaiduApollo} & Camera-LiDAR & Integ. Semantic\\
  \bottomrule
  \end{tabular}
\vspace{-6pt}
\end{table}

\subsection{Attacks on Perception}
\label{sec:att-perception}

\paragraph{Attacks on camera-based perception.}
Camera-based perception algorithms that use DNN models have been shown vulnerable to black-box attacks (e.g.,~\cite{Papernot2017b}). Attacks on camera-based perception have been extended to AV-specific contexts~\cite{Eykholt2018b, Boloor2020b}, showing that object detection and classification are vulnerable when using only camera data.

\paragraph{Demonstrations of LiDAR spoofing attacks.}
Recently, \cite{Petit2015c, Shin2017d, Cao2019d, Sun2020h, Cao2021AutomatedTargets} have demonstrated feasibility of LiDAR spoofing devices. A relay system where LiDAR pulses were received by a photodiode and relayed through an attack laser was introduced in~\cite{Petit2015c}; the system was expanded to control the 3D positioning of spoof points with a delay component~\cite{Shin2017d}, capitalizing on the regular patterning of LiDAR emissions. With this foundation,~\cite{Cao2019d} established a 60 point stable spoofing baseline on a per-frame basis, subsequently~improved to 200 points~\cite{Sun2020h}.

\vspace{-2pt}
\paragraph{Attacks on point cloud detection.}
Spoofing attacks have motivated security studies of LiDAR-based perception. The placement of spoof points is considered as a white-box optimization problem in~\cite{Cao2019d}. In~\cite{Sun2020h}, black-box attacks are introduced, exploiting that DNNs may not encode causality about the data (e.g.,~occluded objects). To date, only mild success is seen in obtaining FPs with spoofed points from a real laser due to engineering limitations~\cite{Sun2020h, Cao2019d}; thus, many security studies use simulated spoofing models~\cite{Sun2020h, Cao2019d, Liu2021} while engineering is improved~\cite{Cao2021AutomatedTargets}.

Further,~\cite{Cao2021InvisibleAttacks} develops physical adversarial objects capable of compromising sensor fusion using gradient-based shape and texture optimization. The model is an expansion on single-sensor adversarial objects, as both camera and LiDAR perception model gradients are used to update shape of the adversarial object. Physical-adversarial-object approaches, such as ~\cite{Cao2021InvisibleAttacks}, require white-box access to the deployed or highly representative perception models for training offline.
Additionally,~\cite{Tu2021b, Abdelfattah2021} introduce attacks with adversarial patches and physical objects that are optimized for color, shape, and texture. Each attack performs optimization over training data~\cite{Tu2021b, Abdelfattah2021}. It has not been studied whether these attacks can generalize across perception algorithms.

\vspace{-4pt}
\subsection{Perception Defenses}
\vspace{-1pt}
Several defenses have been proposed to counteract LiDAR spoofing attacks, including model-agnostic defenses independent of the perception model (e.g.,~CARLO~\cite{Sun2020h}, ShadowCatcher~\cite{Hau2020Shadow-Catcher:Sensing}) and model-based defenses that fortify the perception architecture (e.g.,~SVF~\cite{Sun2020h}, LIFE\cite{Liu2021}).

\textbf{CARLO}~\cite{Sun2020h} is a detection-centric defense, guarding LiDAR-only perception against naive spoofing in front-near positions. The exploits the intuition that, if there are many LiDAR points appearing to pass through a detected object, the object is likely a false positive (FP).
\textbf{ShadowCatcher}~\cite{Hau2020Shadow-Catcher:Sensing} is a detection-centric defense and uses a similar line of reasoning to CARLO: if a detection has a highly anomalous shadow region -- defined as a high anomaly score using features of the shadow region -- it is likely an FP.
\textbf{SVF}~\cite{Sun2020h} is a model-based defense and guards LiDAR-only perception against naive spoofing by augmenting LiDAR data with a point-wise confidence score from the front-view (FV) under the intuition that naive FPs do not maintain FV consistency.
\textbf{LIFE}~\cite{Liu2021} is a hybrid model-based detection-centric perception defense 
that compares LiDAR and camera data detections and raw sensor data. To cross-check sensor detections, the object matching method compares camera and LiDAR detections in the front view. To compare raw sensor data, the corresponding point method checks consistency of camera feature points with raw LiDAR data in a depth image, and the sensor reliability evaluation uses machine-learned prediction algorithms to compare predicted and captured sensor data. LIFE was tested against naive spoofing attacks using LiDAR and stereo imagery~\cite{Liu2021}.

\vspace{-2pt}
\paragraph{Sensor Fusion.} The use of multi-sensor fusion to enhance perception resiliency has been suggested~\cite{Sun2020h, Cao2019d, Ivanov2014b, Liu2021, Pajic2014b}. Yet, no systematic evaluation of sensor fusion under spoofing attacks has been performed (e.g., \textbf{LIFE}~\cite{Liu2021} was evaluated using naive spoofing without analysing spoofing performance,~\cite{Cao2021InvisibleAttacks} used optimized physical adversarial objects as threat model). Thus, in this work we thoroughly evaluate the  fusion models' performance under spoofing~attacks.
\section{Attack Objectives and Threat Model}
\label{sec:3-atack-model}
We use the following terminology in describing the attacker goals, capabilities, and strategy. By the \textit{\textbf{victim}} vehicle, we refer to the AV running perception algorithms. The attack's goal is to cause adverse outcomes for the victim. The attacker may wish to orchestrate attacks in some relation to an object in the scene (e.g.,~another vehicle) other than the victim. This object is referred to as a \textbf{\textit{target}} vehicle. Any other vehicle or object in the scene is denoted as \textbf{\textit{other}}.

\subsection{Attack Goals}
\label{sec:3-attack-model-attack-goals}
We consider false positive (FP) and false negative (FN) attack outcomes consistent with the literature~\cite{Cao2019d, Cao2021InvisibleAttacks, Sun2020h, Tu2021b}, as well as \textit{translation attack outcomes} where a detected object's bounding box is translated (i.e., moved) by some~distance.

\vspace{2pt}
\noindent The goal of achieving a \emph{\textbf{false positive outcome}} is to force the victim to perform dangerous maneuvers (e.g.,~emergency braking or lane change) to avoid the false object. For example, LiDAR spoofing attacks can result in safety-critical incidents, as shown with Baidu's Apollo~\cite{2021BaiduApollo,Sun2020h}.

\vspace{2pt}
\noindent The goal of achieving a \emph{\textbf{false negative outcome}} is to remove an existing object from the perception output such that path planning and control are compromised. Such attacks can have 
the devastating consequence of the victim crashing~into an unsuspecting object hidden to perception (e.g., as in~\cite{Cao2021InvisibleAttacks}).

\vspace{-2pt}
\paragraph{Translation outcome.} We find FP and FN outcomes are insufficient to fully capture the effects of perception attacks. Some cascaded semantic fusion architectures (e.g.,~FPN) enforce one-to-one matching between 2D and 3D detections; thus, an FP necessarily implies an FN. We call such instances \emph{translation outcomes} as the attacker has created physical distance between the negated ground truth (FN) and the spoofed detection (FP). Translation outcomes may cause emergency braking if objects are moved to front-near positions or collision 
when moved farther from the victim or to a different~lane.

\subsection{Threat Model}
\label{sec:3-attack-model-threat-model}

\subsubsection{Environment}
We consider 
scenarios where the victim AV may have multiple sensors;  i.e., we consider both LiDAR-only and camera-LiDAR perception models, 
widely used in AVs~\cite{2021BaiduApollo, 2021NVIDIADRIVE, Hawkins2018, Kato2018AutowareSystems}. 

\subsubsection{Attacker Capability}
We assume the attacker has no access to the AV's internal processing, has no way to attack the camera, and can only inject signal along the same physical channels as normal LiDAR. The attacker uses a LiDAR spoofing attack similar to~\cite{Cao2019d, Sun2020h, Petit2015c, Shin2017d, Cao2021AutomatedTargets}, which established how to control the 3D positioning of LiDAR points using a relay and delay system. Further, we follow the threat model from~\cite{Sun2020h} which demonstrated injecting up to 200 spoof points. While~\cite{Cao2019d, Sun2020h} assume high-precision spoofing where LiDAR points are placed in well-crafted patterns (e.g., outline of a car), we also relax this assumption in some cases by allowing the attacker to place points by randomly sampling a distribution;~the parameters of this distribution are summarized in Table~\ref{tab:point-params}. This simplifies the attack design compared to the model from~\cite{Cao2019d,Sun2020h} and may be more representative of a noisy attack laser.

\begin{table}[!t]
  \caption{\small Gaussian moments for sampling spoof point positions relative to the desired FP location. Coordinate frame is local-level Cartesian, axes are frustum-relative (forward is toward the~victim).}
\vspace{-11pt}
  \centering
  \label{tab:point-params}
  \begin{tabular}{cccc}
    \toprule
    Direction & Forward & Left & Up\\
    \midrule
    Mean (m) & 1.0 & 0 & 1.0\\
    Std. Dev. (m) & 0.1 & 0.5 & 0.2\\
  \bottomrule
\end{tabular}
\vspace{-6pt}
\end{table}

\subsubsection{Attack Strategy}
\label{sec:3-attack-model-attack-strategy}

In this work, we consider the following attack strategies.
\vspace{-2pt}
\paragraph{Naive attacks.} In general, naive attacks compromise a single sensor without regard for consistency between multiple sensors or the environment. Naive LiDAR spoofing attacks against AVs were first proposed in~\cite{Cao2019d} and followed up with~\cite{Sun2020h}. Naive spoofing attacks are examined in Section~\ref{sec:4-naive-spoofing}.

\begin{figure}[!t]
    \centering
    \includegraphics[width=0.68\linewidth]{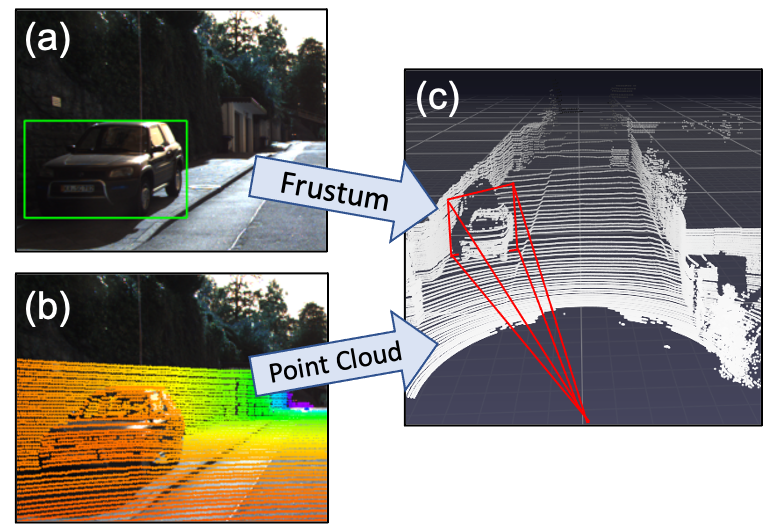}
    \vspace{-8pt}
    \caption{The frustum attack leverages that the camera is only a 2D projection of 3D space. Any feature or detection in a single 2D image could be resolved to any distance along the range axis. Thus, spoofing within the frustum retains consistency between camera and LiDAR data association. The frustum is defined by an object's 2D bounding box when extended into 3D (diagram shown is unattacked).}
     \vspace{-6pt}
    \label{fig:frustum-example}
\end{figure}

\vspace{-2pt}
\paragraph{Frustum attacks.} We introduce the novel \emph{frustum attack} which retains consistency across multiple sensors even only attacking a single sensor. It is motivated by the fact that 2D detections of a target vehicle from the victim camera's front-view cannot resolve range, and thus the 3D uncertainty of a 2D (camera) detection defines a \textit{frustum} from the camera image plane in the direction of the target vehicle (see Fig.~\ref{fig:frustum-example}). Attacking within the frustum of a target vehicle retains consistency with semantic and feature information between camera and LiDAR data. Frustum attacks are examined in Section~\ref{sec:6-frustum-attack}.

\subsubsection{Attacker Knowledge}
\paragraph{System.} In all cases, the attacker requires no knowledge of the underpinnings of perception, including the machine learning model and perception architecture. Further, to instantiate the~attacks, the attacker need not have access to existing sensor data, other than what is required in the relay system~\cite{Petit2015c, Shin2017d}.

\paragraph{Environmental.} For the frustum attacks, we assume the attacker knows the approximate position of the target object so as to obtain a frustum region for spoof point placement. In addition to FP outcomes, this also enables FN or translation attack outcomes targeting a particular (valid) object.

\section{Naive LiDAR Spoofing}
\label{sec:4-naive-spoofing}

We first consider general black-box (i.e., ~naive) attacks on LiDAR-only perception. To fully evaluate the state of the art, we reproduce the LiDAR spoofing attack from~\cite{Sun2020h} using patterns of occluded vehicles extracted from KITTI and sweep number of attack points in steps of 10, from 10~to~200. This is a \textbf{\emph{naive}} method as it does not attempt to maintain consistency among 
sensors and is \textbf{\emph{black-box}} as it does not require knowledge of the employed perception model or sensor data.

\subsection{Spoofing Against LiDAR-Only Perception}
We test a perception algorithm from each of the three categories of LiDAR-only perception architectures, consistent with~\cite{Sun2020h} and outlined in Table~\ref{tab:perception-algs}. Specifically, we use voxel-based PointPillars~\cite{Lang2019a}, point-based PointRCNN~\cite{Shi2019}, and BEV-based PIXOR~\cite{Yang2018a} for 3D object detection. We reproduce the attack success rate (ASR) from~\cite{Sun2020h}. Details on the reproduced results are in Appendix~\ref{sec:appendix-reproduce-attacks}, showing high ASR of the naive spoof attacks at front-near positions.

\subsubsection{State-of-the-Art Defenses}
\label{sec:5-naive-spoofing-defenses}

We reproduced CARLO, SVF, and ShadowCatcher, as no source code was available; reproduced results are presented in Appendix~\ref{sec:appendix-reproduce-defenses}. Our results for CARLO and SVF against naive attacks are consistent with~\cite{Sun2020h} -- i.e.,~the ASR is greatly reduced in front-near positions {against naive attacks}. However, with realistic assumptions, we obtained lower defense performance for ShadowCatcher than reported in~\cite{Hau2020Shadow-Catcher:Sensing}. The reasons, outlined in Appendix~\ref{sec:appendix-reproduce-defenses}, include that the original work tuned parameters on the test set as well as used the ground-truth information instead of the output of a perception algorithm; ground-truth information is not available for a real system and significantly reduces noise.

Very recently,~\cite{Liu2021} introduced LIFE defense that designed point-based and frame-based camera-LiDAR consistency checks as a preprocessing step to guard against both camera and LiDAR attacks. As reported in~\cite{Liu2021}, LIFE is well-suited to detect naive spoofing attacks, as naive spoofing does not retain consistency between the camera and LiDAR~data.

\subsubsection{Some Existing Defenses Have Vulnerabilities}
Under further scrutiny of existing defenses, we find several naive attack configurations not tested in~\cite{Sun2020h} that suggest the CARLO defense introduces additional vulnerabilities.

\begin{figure}[!t]
    \centering
    \includegraphics[width=0.8\linewidth]{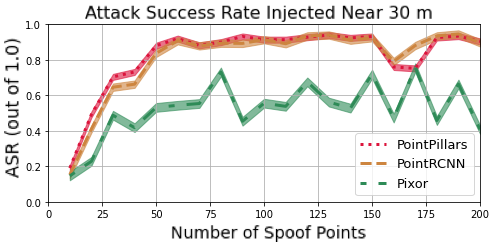}
    \vspace{-10pt}
    \caption{\textit{Naive spoofing attacks against LiDAR-only with CARLO defense outside front-near:} CARLO is not suitable at guarding perception against naive spoofing for false positives outside of front-near; the ASR of CARLO-guarded models is nearly as high as without CARLO (see Fig.~\ref{fig:naive-spoof-8m}) }
    \vspace{-10pt}
    \label{fig:naive-carlo-30m}
\end{figure}

\paragraph{CARLO Vulnerability to False Positives.}
While CARLO demonstrates high success guarding against naive spoofing in front-near~\cite{Sun2020h}, \emph{naive spoofing is stealthy to CARLO when placed outside of front near}. Intuitively, as the range to spoofed objects \emph{increases}, the angle subtended by the frustum towards the detection \emph{decreases}. This leads to a decrease in the number of LiDAR points contained in the frustum as the emitted LiDAR points spread in a spherical pattern (constant angular density). Thus, an increase in range leads to spoofed instances that appear more similar to normal instances under the CARLO hypothesis. We show this by spoofing points at a range of 30~m from the victim; our results in Fig.~\ref{fig:naive-carlo-30m} show that CARLO is incapable of guarding against these naive spoofing attacks, as the ASR is on-par with the defense-less system (compare Figs.~\ref{fig:naive-carlo-30m} and~\ref{fig:naive-carlo-8m}). Analysis for additional spoofed point distances is provided in Appendix~\ref{sec:appendix-defenses-vulnerable}.

\begin{figure*}[!t]
    \centering
    \begin{subfigure}[t]{.24\linewidth}
      \centering
      \includegraphics[width=0.6\textwidth]{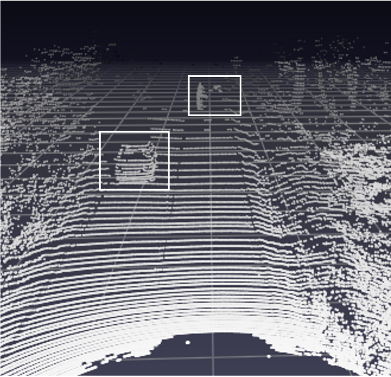} 
      \vspace{-4pt}
      \caption{Stealthy FP at 40 m}
    \end{subfigure}
    \begin{subfigure}[t]{.24\linewidth}
      \centering
      \includegraphics[width=0.6\textwidth]{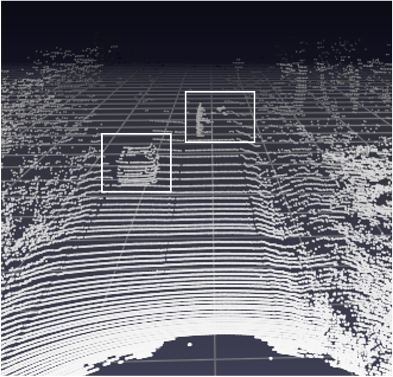}  
      \vspace{-4pt}
      \caption{Stealthy FP at 30 m}
    \end{subfigure}
       \begin{subfigure}[t]{.24\linewidth}
      \centering
      \includegraphics[width=0.6\textwidth]{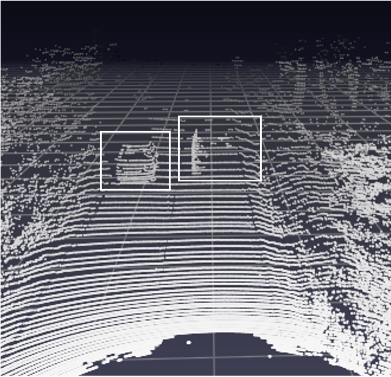}  
      \vspace{-4pt}
      \caption{Stealthy FP at 20 m}
    \end{subfigure}
    \begin{subfigure}[t]{.24\linewidth}
      \centering
      \includegraphics[width=0.6\textwidth]{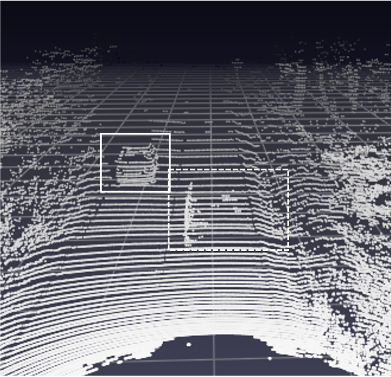}  
      \vspace{-4pt}
      \caption{FP invalidated at 10 m}
    \end{subfigure}
    \vspace{-8pt}
    \caption{Even with the CARLO defense, a spoofing scenario starting at long-range can evade the defense for many frames until it reaches front-near position. During this time, the AV will build a (adversarial) track on the spoofed object, which can cause adverse control outcomes (e.g., collision avoidance maneuver).}
    \vspace{-10pt}
    \label{fig:multi-frame-carlo-fp-case-3d}
\end{figure*}

Spoofing attacks applied at greater ranges \emph{can} have severe adverse outcomes when exercised longitudinally (i.e., over multiple time steps). For example, since LiDAR-only perception cannot be guarded by CARLO outside of front-near, as shown in Fig.~\ref{fig:multi-frame-carlo-fp-case-3d} an adversary can create false positives over multiple time steps to give the appearance of a vehicle moving directly toward the victim with high velocity. This may trigger braking and collision avoidance~maneuvers even before the false vehicle reaches close range.~Detailed investigation of longitudinal attacks is provided in Section~\ref{sec:7-longitudinal}.

\paragraph{CARLO Vulnerability to False Negatives.}
We also find that CARLO, even at front-near, is vulnerable to FN invalidation attacks. Since CARLO relies on physics violations -- i.e.,~many points appearing to ``pass through" a detected object, which should not occur for normal objects -- the attacker can instead use the spoofer to create these physics violations on \emph{normal instances} to obtain FNs (i.e.,~invalidation of true objects). Specifically, an adversary can spoof points behind valid objects (in no particular spoofing pattern, unlike the attacks required by~\cite{Cao2019d, Sun2020h}) which will trigger CARLO into believing the detected object is invalid. Thus, a true object will be rendered as a false negative, potentially causing a head-on collision of the victim vehicle. 

The FN outcomes against CARLO depend on the range to the target to invalidate. This follows from the decrease in the frustum angle with range and the constant angular density of LiDAR points. Using a 200-point maximum capability and requiring only random injections, attacks achieve upwards of $40\%$ success at invalidating objects $50~m$ away (Appendix~\ref{sec:appendix-defenses-vulnerable}).

\begin{figure}[!t]
    \centering
    \includegraphics[width=0.8\linewidth]{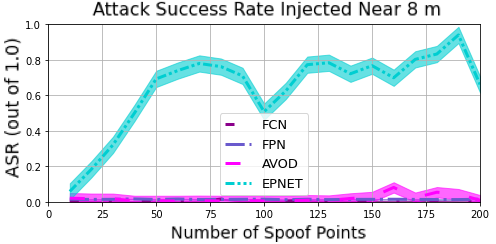}
    \vspace{-10pt}
    \caption{\textit{Naive spoofing attacks against camera-LiDAR fusion:} Many fusion algorithms (including FPN, FCN, AVOD) have high inherent resiliency to naive spoofing attacks ($<5\%$ ASR) without specialized defenses. EPNET's vulnerability is due to its high baseline false positive rate - nearly 40\% of all EPNET's detections are FPs, even without attack.}
    \label{fig:naive-fusion-8m}
    \vspace{-8pt}
\end{figure}

\subsection{Sensor Fusion}
\label{sec:5-naive-spoofing-fusion-defense}
No prior work has systematically evaluated whether sensor fusion is more resilient to spoofing. Thus, we evaluate the naive spoofing attacks against multiple camera-LiDAR fusion algorithms across multiple architectures summarized in Table~\ref{tab:perception-algs}. We find the majority of tested sensor fusion are inherently resilient to the naive spoofing attacks (Fig.~\ref{fig:naive-fusion-8m}). Overall, this level of intrinsic defense renders naive spoofing attacks ineffective even without the addition of specialized~defenses. Specifically, widely used camera-LiDAR fusion algorithms FPN~\cite{Qi2018b}, FCN~\cite{Wang2019c}, and AVOD~\cite{Ku2018} have high resiliency with ASR generally lower than $5\%$. We find EPNET~\cite{Huang2020Epnet:Detection} is still vulnerable; we believe this is due to EPNET's high baseline FP rate. 
On the (unattacked) KITTI validation set, EPNET has 220\% the number of FPs compared to AVOD; nearly 40\% of all EPNET's detections on (unattacked) KITTI~are~FPs. 

\paragraph{Summary: Impact of Black-box Attacks.}
LiDAR-only perception alone is vulnerable to naive black-box spoofing attacks in front-near positions, as previously reported~\cite{Sun2020h, Cao2019d}. However, there are several promising specialized defenses, although CARLO is insufficient in preventing black box spoofing attacks outside front-near positions and is vulnerable to invalidation attacks. Finally, we showed that sensor fusion is intrinsically more robust to naive attacks. Yet, in what follows, we demonstrate that the perception models and defenses perform poorly under a new class of attacks: the \emph{frustum attacks}.
\vspace{-14pt}
\section{Frustum Attack on LiDAR}
\label{sec:6-frustum-attack}
\vspace{-1pt}

In this section, after establishing the feasibility of the frustum attack, we evaluate the impact of frustum spoofing on modern perception methods. We show both LiDAR-only and camera-LiDAR fusion perception are widely vulnerable to the context-aware frustum attacks: \emph{all 8 tested algorithms  falling across 6 different architectures from Table~\ref{tab:perception-algs} are vulnerable} and none of the state-of-the-art defenses against LiDAR spoofing are capable of defending against the frustum~attack.

\subsection{Frustum Attack Motivation \& Definition}
\label{sec:6-frustum-attack-motivation}

While naive spoofing is damaging against LiDAR-only perception, it does not maintain consistency with physical invariants or between camera and LiDAR data; as shown in Section~\ref{sec:4-naive-spoofing}, this inconsistency can be leveraged to filter out the naive spoofing attacks.
Consequently, the \emph{frustum attack} is conceived as a black-box method of retaining consistency between the camera and LiDAR data \emph{and} consistency with physical invariants using easily obtained contextualizing information from the environment. Specifically, an adversary can leverage that the camera is only a 2D projection of the 3D space; any detection or feature in the camera can be resolved to any point along the line extending from the camera out to infinite range (in practice, $\sim 100~m$ for AV applications) because a single camera cannot resolve range information.

The frustum attack thus places spoof points to leverage the projective nature of the camera. Points are placed behind or in front of existing objects so that they have front-view consistency. This can be realized by spoofing within a pyramid (i.e., the \emph{frustum}) where the tip of the pyramid is at the victim sensor and the base is the projection of the 2D, front-view bounding box of a true object out along the range axis.

Thus, due to the projection and by spoofing in-view of existing (target) objects, perception algorithms may associate (unattacked) features/detections in the camera and the spoofed LiDAR points even if the spoof LiDAR points are at a \emph{different} range than the target object. By spoofing in the frustum of valid objects, frustum attack FPs maintain many natural qualities of normal objects (see Fig.~\ref{fig:FPN-Vulnerable}), helping them to be stealthy against existing defenses relying on physical invariants or camera-LiDAR consistency checks.

We denote this `in-view' spoofing as the \emph{\textbf{frustum attack}} since a 2D bounding box around an object in the camera's image defines a frustum when the uncertainty of the 2D box is extended into 3D along the range axis, as illustrated in Fig.~\ref{fig:frustum-example}; also, in the bird's eye view (BEV) in Fig.~\ref{fig:FPN-Vulnerable}.~Importantly, the adversary needs to only approximately know the~frustum.

\vspace{-1pt}
\subsection{Attack Feasibility and Practicality}
\label{sec:3-attack-model-lidar-spoofing}

Feasibility of naive spoofing attacks has been shown in~\cite{Cao2019d, Sun2020h}. Here, we provide experimental justification for the frustum attack. We first describe five situations that naturally arise in nearly all day-to-day driving conditions that enable frustum-based spoofing. We then demonstrate one scenario of the frustum attack experimentally. Three of the situations are attainable with current engineering/LiDAR technology. Work is underway to advance optics and tracking which may enable additional spoofing scenarios (e.g., see~\cite{Cao2021AutomatedTargets}). 

\vspace{-2pt}
\subsubsection{Frustum-Attack Spoofing Scenarios}
We describe five common scenarios where the frustum attack can be exercised; the scenarios are illustrated in Fig.~\ref{fig:spoof-scenarios} in Appendix~\ref{sec:appendix-spoofing-feasibility}. In all cases, the spoofer has full control over the range of placement of the spoof points along the frustum by increasing or decreasing the delay timing. 

\vspace{-3pt}
\paragraph{S1: Spoofer on target.} A spoofer is placed on a target car and aimed at victim (the target AV owner/passengers may or may not be aware of this). The target car does not have to be endangered for this to have impact because the attacker creates FPs that cause the victim to perform evasive maneuvers. The target car is by definition in line with its own frustum. Any spoofed points along the line-of-sight (LOS) between the target and the victim will remain in the frustum.
\vspace{-3pt}
\paragraph{S2: Spoofer on other vehicle in line.} Spoofer is placed on a non-target car on the line defined by the victim AV and target vehicle. This scenario arises often in natural driving, as a lane, which is usually locally straight, helps cars stay in line with each other, and thus in each others frustums.
\vspace{-3pt}
\paragraph{S3: Spoofer on other vehicle not in line.} A fully general spoofing attack could take place out-of-line. However, executing this attack outside the frustum is not currently feasible and requires more precise aiming of the laser than has been demonstrated. Engineering advances will enable this scenario, and work is already underway in this area~\cite{Cao2021AutomatedTargets}.
\vspace{-3pt}
\paragraph{S4: Spoofer on environment in line.} A spoofer is placed in the environment in line with a lane. Examples include placing the spoofer on a bridge transverse to the road or on low-lying traffic signs, tree limbs, etc.
\vspace{-3pt}
\paragraph{S5: Spoofer on environment not in line.} This 
resembles \textbf{S3} with similar feasibility constraints but with a spoofer placed on a static object in the environment (e.g., road-side sign).

\subsubsection{Feasibility Demonstration}
We adopt the physical hardware from~\cite{Cao2019d, Sun2020h} and use a VLP-16 PUCK for the LiDAR sensor and for the spoofing system, an OSRAM SFH 213 FA photodiode, an OSRAM SPL PL90 attack laser, and an additional lens for beam focusing. The VLP-16 is a rotating LiDAR scanner providing full $360^{\circ}$ azimuth coverage and is compatible with many modern industry AVs and perception, including LGSVL~\cite{Rong2020LgsvlDriving} and Baidu Apollo~\cite{2021BaiduApollo}, which have LiDAR plugins for the VLP-16.

\begin{figure}[!t]
    \centering
    \includegraphics[width=0.57\linewidth]{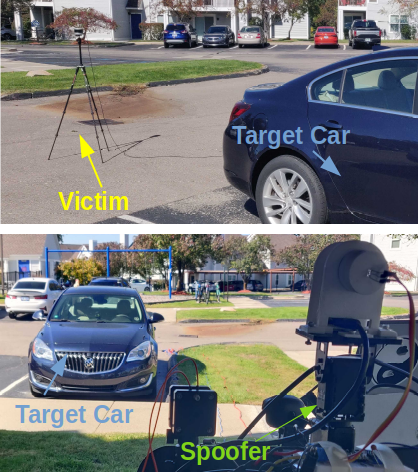}
    \vspace{-8pt}
    \caption{A spoofer launches a malicious \emph{frustum attack} against a victim AV using a target car. Spoof points are placed at any distance within the frustum behind the target car to obtain false positive, false negative, and/or translation outcomes.}
    \vspace{-4pt}
    \label{fig:spoof-diagram-1}
\end{figure}

To test frustum attack feasibility, the spoofing device is placed behind the target vehicle (Fig.~\ref{fig:spoof-diagram-1}). The spoofer has just enough visibility above the target car for the attack laser to have LOS to the LiDAR sensor. This is easily realized in everyday driving so long as the attacking vehicle is slightly larger than the victim or the spoofing device is elevated (e.g.,~placed on the roof of AVs, like existing LiDAR).

\begin{figure}[!t]
    \centering
    \includegraphics[width=0.5\linewidth]{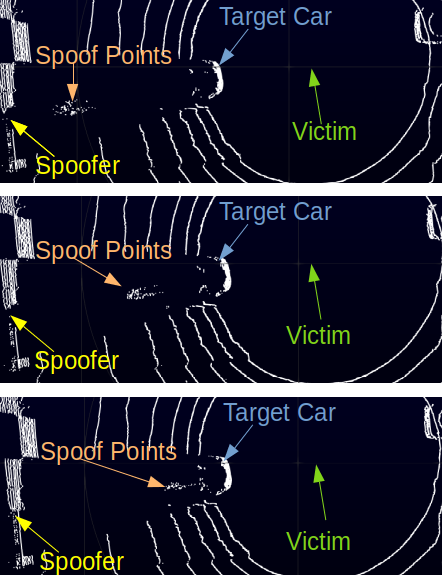}
    \vspace{-6pt}
    \caption{A physical experiment demonstrates that an attacker can stably spoof longitudinally consistent points in the frustum of a target vehicle and has control over the placement distance. We run a continuous scenario where spoof points are moved successively towards and away from the target, and show three select frames here. Running FPN as perception would cause a \emph{translation outcome} where the target car would be detected at the spoof point location. Full video online~\cite{TheAttack}.}
    \vspace{-8pt}
    \label{fig:spoof-diagram-2}
\end{figure}

We find the spoofer can command the delay timing to inject spoof clusters at varying distances relative to the target car. In Fig.~\ref{fig:spoof-diagram-2}, spoof point clusters are moved successively farther from the target car (or closer, if run in reverse) in a dynamic environment with longitudinal consistency. We also repeat the experiment with a moving target car but stationary victim and spoofer; for conciseness, those scenarios (including videos) are only available online, along with the project code, at~\cite{TheAttack}.

\vspace{-2pt}
\paragraph{Discussion.}
The above experiments cover two situations: (1) victim, target, and spoofer are in-line and (relatively) static which encompasses \textbf{S1} and \textbf{S2} for vehicles traveling in unison (e.g., vehicle platooning), and (2) target moving relative to spoofer which encompasses the same prior situations (this time with relative motion) as well as \textbf{S4} due to the relative velocity between the vehicles. The outcomes of these spoofing experiments validate that common, everyday spoofing scenarios are feasible even with existing spoofing hardware, although executing the frustum attack with motion of the spoofer and target has not been fully demonstrated.

In fact, a frustum attack only requires LOS between the spoofing device and the victim AV with at least one object in the scene. The victim and target vehicles always define a frustum, so it is up to the attacker to position the spoof points within the frustum; this is trivially satisfied when the spoofer is in the same lane as the vehicles (e.g.,~on another car) or may simply require a lane change or velocity adjustment. With improvements in laser aiming, the number of natural frustum attack scenarios will only increase~\cite{Cao2021AutomatedTargets}.

Finally, if started near the target vehicle, moving spoof~point clusters, as in videos online~\cite{TheAttack}, can create longitudinally-consistent frustum attacks that drift the track further away (or closer to the victim, if run in reverse), as described in Section~\ref{sec:7-longitudinal-tracking-case-studies}. Case studies of the related tracking and end-to-end control outcomes are presented in Sections~\ref{sec:7-longitudinal-tracking-case-studies} and~\ref{sec:7-longitudinal-baidu-apollo}.

\subsection{Frustum Attack Performance Analysis}

\subsubsection{Experimental Methods}
\label{sec:6-frustum-attack-experiment-details}
We run a large-scale experiment on over 75 million scenarios to assess the vulnerability of perception to the frustum attack for FP, FN, and translation outcomes.

For each of the first 7 perception algorithms in Table~\ref{tab:perception-algs}, we select each valid vehicle in each frame of the KITTI validation set. Each vehicle becomes the "target vehicle". For our analysis, we discard any valid vehicles 
not detected by the unattacked perception algorithm, as this would artificially inflate the FN attack success metric. For each valid vehicle, we simulate frustum attack spoofing using different combinations of the number of spoof points $n_i$ and the relative distance of placement $d_i$, i.e., $(n_i, d_i)$, within $n_i\in[2, 200]$ points and $d_i\in[r_0-10, r_0+30]~m$; here, $r_0$ is the original range to the target. 
The experiment captures existence of spoofing-induced: (a) FP at the spoof location, and (b) FN of the target object. 

This experiment yielded on average 11 million attack traces for each perception algorithm with a total of over 75 million attack traces for the frustum attack. We also assess the four aforementioned (three experimentally, one in discussion) defenses for each perception algorithm. Due to the combinatorial nature of such evaluation (algorithm $\times$ points $\times$ distance $\times$ defense), we sample a set of attack traces over a coarse grid of parameters for each tested perception~algorithm.

\begin{figure}[!t]
    \begin{subfigure}{.45\linewidth}
      \centering
      \includegraphics[width=.8\textwidth]{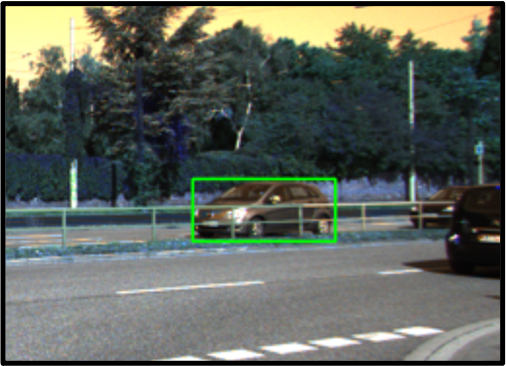}  
      \caption{Clean RGB detection}
    \end{subfigure}
    \begin{subfigure}{.45\linewidth}
      \centering
      \includegraphics[width=.8\textwidth]{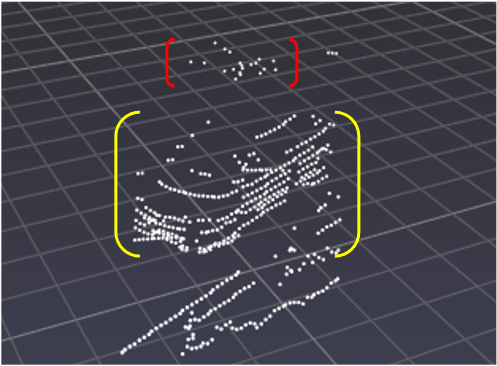}  
      \caption{Frustum with attack}
      \label{fig:FPN-Vulnerable-b}
    \end{subfigure}
    \begin{subfigure}{\linewidth}
      \centering
      \includegraphics[width=.55\textwidth]{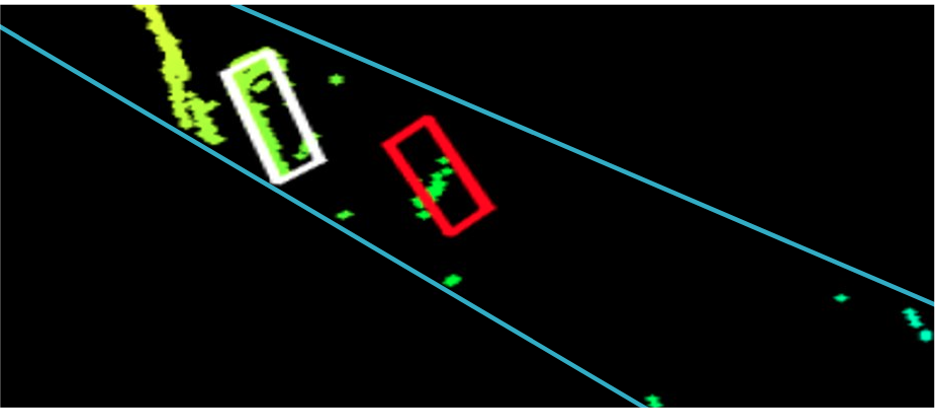}  
      \caption{BEV shows FPN detects inserted points, not original points.}
    \end{subfigure}
    \vspace{-6pt}
    \caption{(a) 2D detection yields (b) a 3D frustum of points.~Injecting just 20 points in random pattern (bracketed red) deceives 3D object detection, even against a valid object (bracketed yellow) of 238 points; (c) BEV projection of the 3D detection show success of the frustum attack with translation outcomes, as the FP detection (red box) is far from the FN ground truth~(white~box).}
    \Description{Frustum pointnet vulnerability example.}
    \vspace{-8pt}
    \label{fig:FPN-Vulnerable}
\end{figure}

\vspace{-2pt}
\paragraph{Example outcome.}
An example successful frustum attack against FPN fusion is in Fig.~\ref{fig:FPN-Vulnerable}, where 20 spoofed points~are placed in a random pattern with a mean location $7~m$ behind~and within the frustum of a target valid object. We find that, as long as spoof points are within the frustum, it is less important how precise those points are placed. In fact, we find in general that spoofing using a normal distribution of points with moments specified in Table~\ref{tab:point-params} can achieve performance on-par with extracting occluded traces from KITTI as done in~\cite{Sun2020h} (see Appendix~\ref{sec:appendix-spoof-placement} for detailed comparison).

In this case, the target object is composed of 238 points, an order of magnitude more than the spoofed points, and is at $25~m$ range from the victim. As shown in Fig.~\ref{fig:FPN-Vulnerable}, even only attacking LiDAR, the frustum attack is successful in obtaining an FP at the spoof point~cluster.

\subsubsection{Results~I: Frustum Attacks Compromise All Perception Algorithms}
\label{sec:6-frustum-attack-results}
We now show that the frustum attack is capable of not only compromising LiDAR-only perception but also compromising camera-LiDAR fusion. Also, the frustum attack succeeds across multiple architectures of both LiDAR-only and camera-LiDAR perception. Here, we describe the main observed results. Additional results are presented in Appendix~\ref{sec:appendix-frustum-attack-all-algorithms}.

\vspace{-4pt}
\subsubsection*{Attackability: Attack Existence}
A frustum is "attackable" if there is at least one combination $(n_i, d_i)$ within the established attacker capability that is successful in generating an FP near the spoof points (or FN of the targeted object, depending on attacker's goal). Given a fixed set of input sensor data, the vulnerability of a perception algorithm depends on how many target objects are attackable.

\begin{figure}[!t]
    \begin{subfigure}[c]{\linewidth}
      \centering
      \includegraphics[width=0.72\textwidth]{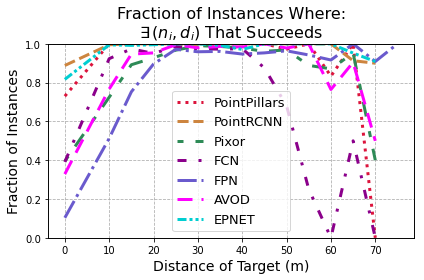}
      \vspace{-6pt}
      \caption{For FP: All algorithms are highly attackable for FP outcomes, particularly when the target objects are at $15-60$~m range -- here, the attackability is near 100\% across the board (except FCN's dip).}
      \label{fig:exists-fp-attack-over-dist-all-pts}
    \end{subfigure}
    \begin{subfigure}[c]{\linewidth}
      \centering
      \includegraphics[width=0.72\textwidth]{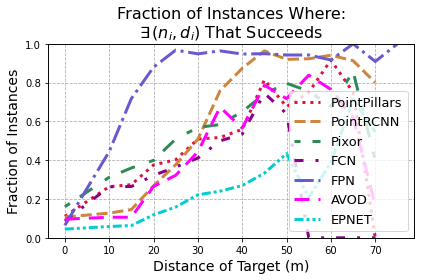}
      \vspace{-6pt}
      \caption{For FN: Perception demonstrates a surprising vulnerability to FN outcomes under LiDAR spoofing. The targeted object can be negated (i.e.,~not detected) for all perception algorithms, even under a small, spoofing frustum attack model.}
      \label{fig:exists-fn-attack-over-dist-all-pts}
    \end{subfigure}
    \vspace{-8pt}
    \caption{Percentage of instances in the KITTI dataset (over number of points and distance of placement) where there exists a successful (a)~FP, and (b) FN frustum attack; all perception algorithms show widespread vulnerability to both (a)~FP and (b)~FN outcomes under the frustum attack. 
    }
    \vspace{-8pt}
    \label{fig:exists-attack-over-dist-all-pts}
\end{figure}

Fig.~\ref{fig:exists-attack-over-dist-all-pts} illustrates the vulnerability of each perception algorithm by presenting the fraction of target objects that are attackable. Presented are both FP (Fig.~\ref{fig:exists-fp-attack-over-dist-all-pts}) and FN (Fig.~\ref{fig:exists-fn-attack-over-dist-all-pts}) outcomes to comprehensively illustrate the vulnerability. Considering FP outcomes, at middle ranges to target objects (i.e., $15-40$~m) \emph{nearly 100\% of instances using any of the perception algorithms are attackable}, showing the widespread vulnerability to frustum attacks. In fact, except for FCN which has a dip in attackability from $40-60$~m, this near-100\% attackability extends for all other perception algorithms from $15-60$~m, which is a devastating outcome for AV perception.

Similarly, we show a surprisingly high degree of FN vulnerability (Fig.~\ref{fig:exists-fn-attack-over-dist-all-pts}) even after discarding targeted vehicles not detected without attack. At a $35$~m range, 
with a suitable selection of spoof distance,
half of all vehicles can be negated with a frustum spoofing attack for all algorithms except EPNET.

Due to space constraints, in the rest of the work, we focus on analysis of FP outcomes which we find are more successful, repeatable, and adaptable to different spoofing distances compared to FN outcomes.

\vspace{-4pt}
\subsubsection*{Attack Success Across Number of Spoofed Points}
Attack success depends on the number of spoofed points, up to a certain point of convergence. Here, we look at two key indicators of spoofing success: 1)~the rate of attack success across the range to target objects for discrete numbers of spoof points, and 2)~the minimum attacker requirements for successful attacks in general. 

\begin{figure}[!t]
    \begin{subfigure}[c]{\linewidth}
      \centering
      \includegraphics[width=0.62\textwidth]{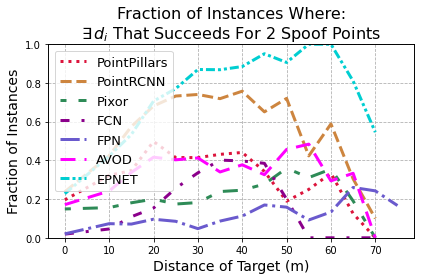}
      \label{fig:exists-fp-attack-over-dist-2-pts}
    \end{subfigure}
    \begin{subfigure}[c]{\linewidth}
      \centering
      \includegraphics[width=0.62\textwidth]{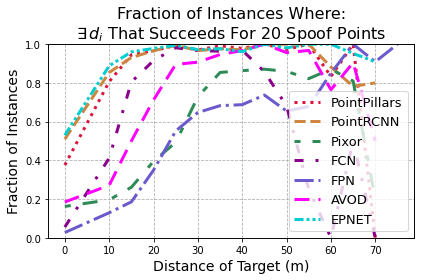}
      \label{fig:exists-fp-attack-over-dist-20-pts}
    \end{subfigure}
    \begin{subfigure}[c]{\linewidth}
      \centering
      \includegraphics[width=0.62\textwidth]{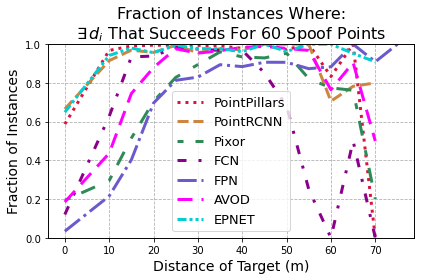}
      \label{fig:exists-fp-attack-over-dist-60-pts}
    \end{subfigure}
    \vspace{-10pt}
    \caption{Percentage of instances in the KITTI dataset (over the placement distance) where there exists a successful FP attack for different numbers of spoof point: the frustum attacks are successful across a wide 
     range of spoof~point numbers. 
    }
    \vspace{-10pt}
    \label{fig:exists-fp-attack-over-dist-by-points}
\end{figure}

\vspace{-4pt}
\paragraph{Attack success by number of points.}
Fig.~\ref{fig:exists-fp-attack-over-dist-by-points} presents how attack success depends on the discrete numbers of spoofed points. Surprisingly, even spoofing just 2 points may be enough to obtain FP outcomes at the site of spoofed point placement given an optimal selection of spoof point distance. The attack success quickly converges to a rate similar to the one in Fig.~\ref{fig:exists-attack-over-dist-all-pts} at just 60 spoof points. 

\begin{figure}[!t]
    \centering
    \includegraphics[width=0.62\linewidth]{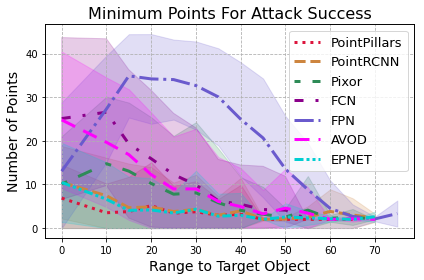}
    \vspace{-10pt}
    \caption{For each perception algorithm, the mean of the smallest number of spoof points for a successful FP attack, marginalized over the relative spoof placement, is low. Generally, the mean smallest set is less than 20 for nearly all algorithms for all ranges to target vehicles.}
    \label{fig:minimum-points-fp-asr}
    \vspace{-6pt}
\end{figure}

\vspace{-4pt}
\paragraph{Minimum attack requirement.} In general, more spoof points yields higher ASR. However, since an attacker only needs a few successful attacks to cause devastating outcomes and may not have the ability to spoof large numbers (e.g.,~hundreds) of spoof points, it is important to understand the average smallest number of points needed for a successful attack. To compute this estimate of the 0th order statistic of spoofing, for each perception algorithm, if the target vehicle were attackable, we logged the range to that target object, $r_0$, and stored the smallest number of points, $n_{i,\min}$, where an attack succeeded, marginalizing over distance, $d_i$. We then computed the mean of this collection of minima against range to the target object (see Fig.~\ref{fig:minimum-points-fp-asr}) and find that only tens of points are needed on average.
Note that these results can be interpreted as a measure of robustness of the perception algorithm to small numbers of spoof points; e.g., FPN is significantly more robust at intermediate ranges, FCN, PIXOR, and AVOD are more robust than PointPillars, PointRCNN, and EPNET.

\vspace{-4pt}
\subsubsection*{Attack Success Across Range to Target Vehicle}
\begin{figure*}[t!]
\begin{subfigure}[t]{0.24\linewidth}
  \centering
  \includegraphics[width=.9\textwidth]{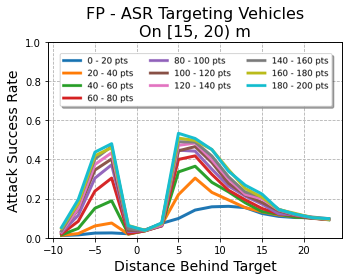}
  \vspace{-4pt}
  \caption{FP ASR on vehicles [15, 20)m}
\end{subfigure}
\begin{subfigure}[t]{0.24\linewidth}
  \centering
  \includegraphics[width=0.9\textwidth]{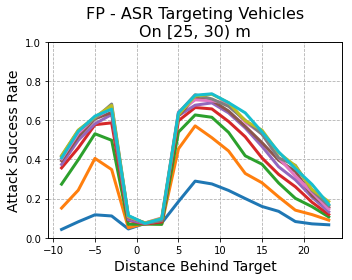}
  \vspace{-4pt}
  \caption{FP ASR on vehicles [25, 30)m}
\end{subfigure}
\begin{subfigure}[t]{0.24\linewidth}
  \centering
  \includegraphics[width=0.9\textwidth]{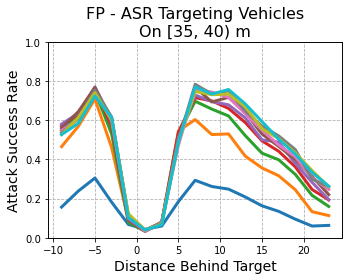}
  \vspace{-4pt}
  \caption{FP ASR on vehicles [35, 40)m}
\end{subfigure}
\begin{subfigure}[t]{0.24\linewidth}
  \centering
  \includegraphics[width=0.9\textwidth]{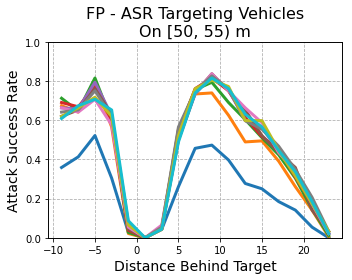}
  \vspace{-4pt}
  \caption{FP ASR on vehicles [50, 55)m}
\end{subfigure}
\vspace{-6pt}
\caption{ASR when AVOD is used as a function of the spoof points' distance (relative to the target vehicle). Attacks are more successful at increased range of the target ((a) vs. (b), (c), (d)). Horizontal axis represents relative placement of spoof~points;~each line represents a different number of spoof points from $0-200$. Number of points determines attack success up to a steady~state where additional points provide marginal benefit. High FPs 
are seen spoofing both in front (-x axis) and behind (+x) the~target.}
\vspace{-6pt}
\label{fig:attack-dist-avod}
\end{figure*}

The location of the target car is an important element~in~the success of a frustum attack. For different ranges to target~vehicles, Fig.~\ref{fig:attack-dist-avod} breaks down the success against AVOD as a function of both the placement of the spoof points relative to the target vehicle and the number of spoof points and does not marginalize over parameters. This highlights that spoofing attacks are generally more successful as the range to target vehicles increases. A similar pattern is observed across all tested perception algorithms, as shown in Fig.~\ref{fig:attack-dist-all-algorithms} in Appendix~\ref{sec:appendix-frustum-attack-all-algorithms}.

\subsubsection{Results II: Frustum Attacks Compromise Defenses}
We show that in addition to being effective against both LiDAR-only and camera-LiDAR fusion, the frustum attack is stealthy to the aforementioned defenses.

We collected a sample of attack traces using each pairwise combination of spoof points in $n_i\in\{10, 60, 100, 200\}$ and attack distance $d_i\in r_0+\{5, 9, 12, 16\}~m$ and run each combination for 200 frames of data for each algorithm, totaling nearly 25,000 attack traces per each defense. We observed that stealthiness to the defenses as a function of the parameters is nearly constant; thus, aggregated results across parameters are summarized in Tables~\ref{tab:frustum-against-carlo},~\ref{tab:frustum-against-svf},~\ref{tab:frustum-against-gbsc}. We report the fraction of frustum attacks that are still successful after applying the defense as the "\% Stealthy", and, where relevant, 
the fraction of erroneously invalidated valid objects as the "Induced~FN~Rate".

\vspace{-2pt}
\paragraph{CARLO:} The frustum attack against all perception algorithms is \emph{nearly completely stealthy} to the CARLO defense since attacks are placed in the frustum and few LiDAR points travel through the spoofed object; see results in Table~\ref{tab:frustum-against-carlo}. 

\begin{table}
  \caption{Nearly all frustum attacks against both fusion (left) and LiDAR-only (right) are stealthy to CARLO defense}
  \centering
  \label{tab:frustum-against-carlo}
  \vspace{-10pt}
  \begin{tabular}[t]{cc}
    \toprule
    Algorithm & \% Stealthy\\
    \midrule
    FCN & 100\%\\
    FPN & 99.76\%\\
    AVOD & 100\%\\
    EPNET & 99.9\%\\
  \bottomrule
  \end{tabular}
  \begin{tabular}[t]{cc}
    \toprule
    Algorithm & \% Stealthy\\
    \midrule
    PointPillars & 100\%\\
    PointRCNN & 99.9\%\\
    PIXOR & 92.3\%\\
  \bottomrule
  \end{tabular}
\vspace{-4pt}
\end{table}

\vspace{-2pt}
\paragraph{SVF:} The frustum attack is stealthy to the SVF defense since frustum spoofs are consistent with information from the front-view projection; see results in Table~\ref{tab:frustum-against-svf}.

\begin{table}
  \caption{Frustum attack is stealthy to SVF defense}
\vspace{-11pt}
  \centering
  \label{tab:frustum-against-svf}
  \begin{tabular}[t]{cc}
    \toprule
    Algorithm & \% Stealthy\\
    \midrule
    SVF-PointPillars & 90.3\%\\
  \bottomrule
  \end{tabular}
\vspace{-4pt}
\end{table}

\vspace{-2pt}
\paragraph{Shadow-Catcher:} ShadowCatcher does not perform well at detecting the frustum attack or identifying normal objects as valid, as confirmed by results in Table~\ref{tab:frustum-against-gbsc}. Our results show an unacceptably high induced FN rate (i.e.,~it invalidates true objects at too high of a rate).

\begin{table} [!t]
  \caption{ShadowCatcher fails to detect a significant number of frustum attacks and has too high induced FN rate.}
\vspace{-11pt}
  \centering
  \label{tab:frustum-against-gbsc}
  \begin{tabular}[t]{ccc}
    \toprule
    Algorithm & \% Stealthy & \% Induced FN Rate\\
    \midrule
    FCN & 80.7\% & 70.3\%\\
    FPN & 57.8\% & 96.9\%\\
    AVOD & 84.9\% & 72.9\%\\
    EPNET & 90.5\% & 64.4\%\\
  \bottomrule
  \end{tabular}
  \vspace{4pt}
  \begin{tabular}[t]{ccc}
    \toprule
    Algorithm & \% Stealthy & \% Induced FN Rate\\
    \midrule
    PointPillars & 91.0\% & 68.0\%\\
    PointRCNN & 89.3\% & 67.1\%\\
    PIXOR & 81.5\% & 42.9\%\\
  \bottomrule
  \end{tabular}
\vspace{-10pt}
\end{table}

\vspace{-2pt}
\paragraph{LIFE:}
LIFE is designed to identify faults, miscalibrations, and attacks against AVs equipped with panoptic stereo cameras and a central, wide-angle or scanning LiDAR sensor using an Object Matching Method (OMM), a corresponding point method (CPM), and a sensor reliability evaluation (SREM)~\cite{Liu2021}. 
However, each of these components are ill-posed for detecting the frustum attack, as noted even by the authors in~\cite{Liu2021}; specifically, Section~8.4.1 of~\cite{Liu2021} states that a common failure mode is when "\emph{most injected fake echoes/points are behind or very near existing aboveground objects...the induced fake objects cannot be detected.}" Specifically, OMM fails to detect the frustum attack because it uses a projection of LiDAR onto the 2D image plane to check consistency between 2D image and 2D LiDAR -- the frustum attack is designed to retain consistency for this very purpose. Second, CPM fails to detect the frustum attack because it generates a small set of 3D features from the camera, \textit{then} checks for a corresponding LiDAR point. Thus, CPM \textit{cannot detect the frustum attack} as 
it maintains consistency with the camera data and is placed in sparse regions where no checking will occur. Finally, SREM projects LiDAR to the image plane and compares the 2D camera and 2D (front-view) projected LiDAR where the frustum attack is designed to be consistent. 


\subsubsection{Security Implications}
The presented results establish that the \textit{frustum attack} is successful in compromising both LiDAR-only perception as well as camera-LiDAR fusion, whereas existing state-of-the-art defenses against LiDAR spoofing are ineffective against the frustum attack. Consequently, \emph{existing perception algorithms are not secure against LiDAR spoofing} when additional contextual information is available for identifying frustums.
\vspace{-2pt}
\section{Longitudinal Case Studies}
\label{sec:7-longitudinal}
Isolated instances of spurious attacks on perception will not survive against real AVs with multiple sensors capturing data over time. With map-aided tracking, AVs can flag FPs that do not comply with semantic map or dynamics information. Tracking also builds resiliency to isolated FNs by allowing for coast time in between measurements~\cite{Blackman1986}.

The frustum attack, with robustness to number of points and distance of injections, as well as success against multiple algorithms and random spoof patterns, is suitable for temporally consistent spoofing to achieve impact at the tracking level (i.e.,~over time). The physical spoofing experiments from Section~\ref{sec:3-attack-model-lidar-spoofing} and linked videos~\cite{TheAttack} show longitudinal frustum attacks where a spoofer can gradually adjust the position of spoof points to simulate motion of a spoofed object. We provide additional visualizations to understand longitudinal frustum attacks in Appendix~\ref{sec:appendix-tracking-case-studies}, Fig.~\ref{fig:frustum-longitudinal-visualization}.

To confirm impact of the frustum attack on real systems using temporal fusion, we perform two evaluations. First, we explicitly analyze spoofing's impact on the multi-frame tracking algorithm and present two case studies showing that such attacks jeopardize AV safety. Second, we apply the frustum attack to an end-to-end, industry-level AV software stack, Baidu Apollo~\cite{2021BaiduApollo} using the LGSVL simulator~\cite{Rong2020LgsvlDriving} and show resulting adverse planning and control outcomes.

 \vspace{-2pt}
\subsection{Frustum Attack Impact on Tracking}
\label{sec:7-longitudinal-tracking-case-studies}

\subsubsection{Tracking Algorithm}
We implement a Kalman filter tracker with position, velocity, and acceleration states according to~\cite{Blackman1986}. All major industry players, including Baidu Apollo~\cite{2021BaiduApollo}, Autoware~\cite{Kato2018AutowareSystems}, and OpenPilot~\cite{CommaAI} use variations on the Kalman Filter for tracking and fusion. We use one tracker per frustum using FPN as perception, as FPN encodes a one-object-per-frustum requirement. The track (i.e.,~trajectory) is predicted forward using a nearly-constant acceleration model and process noise according to~\cite{Li2003}, which is consistent with industry-level AVs~\cite{2021BaiduApollo}. 3D detections from camera-LiDAR perception are fed at 5~Hz to the tracking module which tracks box centers over time. We use an industry-standard $\chi^2$ gating between predicted tracks and timestamped measurements as tracking integrity; this ensures temporal consistency between measurements and prevents unlikely associations from updating tracks. We use the $99\%$ threshold of the $\chi^2$ gate, specified~as $
       0.99 = \text{Pr}(g_k > \tau),  \text{~where~~}
        g_k = z^T Q^{-1} z;
$
here, $\tau$ is the threshold found using the $\chi^2$ inverse CDF, $z$ is the innovation between propagated state and measurement, and $Q$ is the innovation covariance from the Kalman update~\cite{Blackman1986}.
In other words, we neither forced perception to detect our spoofed points nor did we force tracks to accept the resulting detection. We fixed attacker capability at 65 points, which is substantially less than the maximum demonstrated capability. 

\vspace{-2pt}
\subsubsection{Scenario I: Vehicles at Intersection}
We first consider an attacker creating an adversarial track on a crash course for collision with the victim. We select a scene where the target is at $35~m$ range. With traffic lanes $4~m$ wide and vehicles $5~m$ long, this scenario represents a large intersection where the cars are initially static.

Due to perception's high frame rate, the attacker need only to succeed in attacking over a short time window for a false track to be created. The attacker 
injects 10 sequences of point clusters behind the target, corresponding~to $2~s$ of real-time, and alters the distance between successive spoofs so that the vehicle appears to accelerate towards~the~victim.

\begin{figure}[!t]
  \centering
  \includegraphics[width=0.84\linewidth]{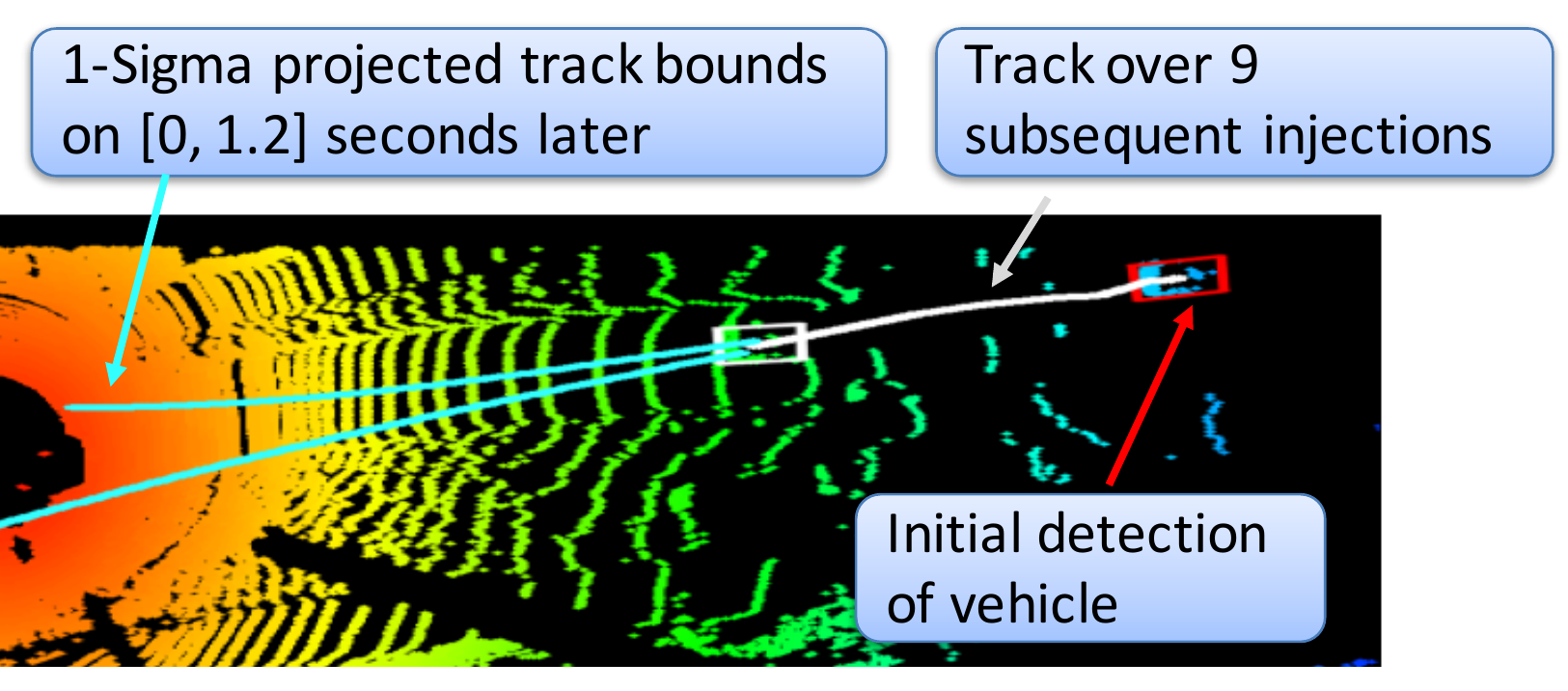}  
  \vspace{-10pt}
  \caption{Attacking over multiple frames at attacker-specified distances creates adversarial tracks. Sequences of 65 point spoofs create false detections that are accepted by tracker integrity. Initial spoof detection (red) travels along (white) track with time-to-impact with the victim vehicle predicted $1$~s later, with high-certainty (cyan) near impact.}
  \Description{Tracking motivation from BEV.}
  \vspace{-6pt}
  \label{fig:tracking-case-1}
\end{figure}

Fig.~\ref{fig:tracking-case-1} illustrates the BEV of the false track created from this spoofing attack. Eight of the attacker's ten injections were falsely detected by perception and accepted by the $\chi^2$ gate to update the created (adversarial) track. 

For path planning, it is essential that AVs understand both the current states of nearby vehicles as well as their future trajectories in order to plan a safe path through the environment. After two seconds of attack, a path planner predicts the existing track forward, shown in Fig.~\ref{fig:tracking-case-1}, and the vehicle in front of the victim is on a collision course with a time-to-impact of just over $1~s$. This can trigger dangerous, aggressive and unnecessary collision avoidance~maneuvers.

\vspace{-2pt}
\subsubsection{Scenario II: Highway Adaptive Cruise Control}
\label{sec:scanario2}
Here, we consider highway flow of traffic (e.g.,~$25$~m/s) where adaptive cruise control uses perception to monitor objects and to keep up with traffic flow. We consider a likely case in which the victim AV has already achieved high-precision track on a true vehicle in front. An existing high-confidence track is more challenging for an attacker to manipulate 
(e.g., see~\cite{Jia2020a}). In this case, any dramatic deviation in the location of that object may trigger an alarm or rejection by the $\chi^2$ integrity monitor, particularly since perception operates~at~high~rate.

\begin{figure}[!t]
  \centering
  \includegraphics[width=0.86\linewidth]{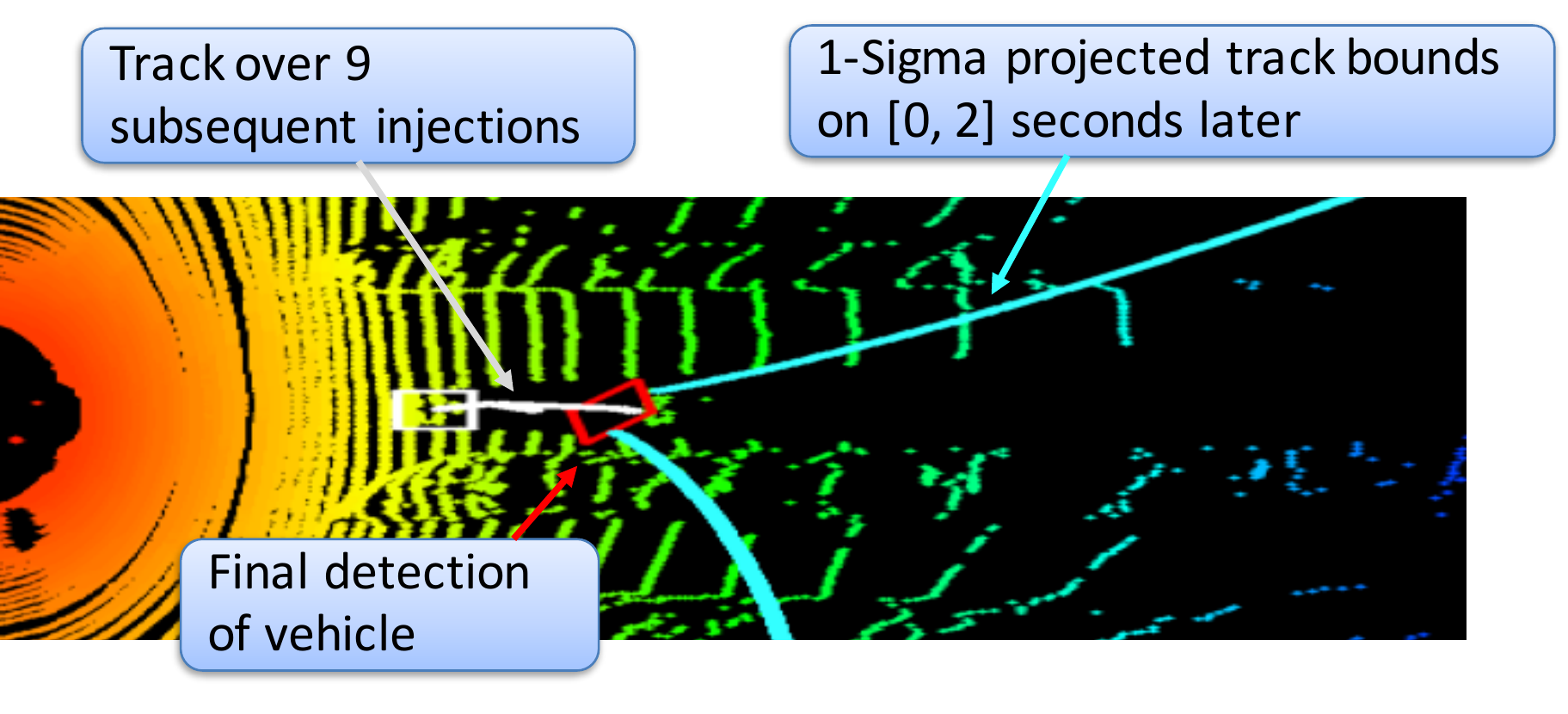}  
  \vspace{-12pt}
  \caption{Attacking 5 frames using only 60 points can transform a high-confidence track of a valid vehicle into a low-confidence track with adversarial velocity away from victim. Predicted path shows target vehicle is moving away when, in reality, the target and victim vehicles have no relative velocity.}
  \vspace{-6pt}
  \label{fig:tracking-case-2}
\end{figure}

Over just five spoofs which corresponds to $1$~s of real-time, an attacker can manipulate an existing track by gradually increasing the distance of spoof points away from the target (Fig.~\ref{fig:tracking-case-2}). While initially the track has no relative velocity (i.e.,~vehicles traveling in unison), path planning updates track prediction for the vehicle in front after the frustum attack, and it appears to travel away at an accelerating rate. This will cause an increase in the adaptive cruise control speed of the victim vehicle, due to the \emph{apparent} increased velocity of the traffic flow, when in reality, the ground truth vehicle is still traveling with no relative velocity; thus, the victim vehicle will dangerously approach the car in~front.

\vspace{-2pt}
\subsection{Frustum Attack Impact At Driving-Decision Level: Baidu Apollo Study}
\label{sec:7-longitudinal-baidu-apollo}
We perform another case study using the high-fidelity physics-based simulator, LGSVL~\cite{Rong2020LgsvlDriving}, and the Baidu Apollo AV stack~\cite{2021BaiduApollo}. We use LGSVL with a 32-beam LiDAR model and a Full-HD 1080p camera model to capture realistic LiDAR and camera sensor data for the frustum attack. While the physics engine of LGSVL is built on Unity and robust, the LGSVL API is under continuous development. As a result, it is unclear how to modify low-level sensor data in real-time. Thus, we took a multi-stage approach to evaluating the end-to-end consequences of the frustum attack on Apollo's control. First, we capture LiDAR and camera data during normal operation. Second, we execute the frustum attack on the captured data and run Apollo offline to get detected objects and control decisions. Finally, we replay the control data through the LGSVL bridge to observe and visualize the outcomes. We were able to use this approach as the control commands of the vehicles were matching in the first and second runs, up to the point when the victim vehicle initiates emergency braking.

\begin{figure}
    \centering
    \includegraphics[width=.95\linewidth]{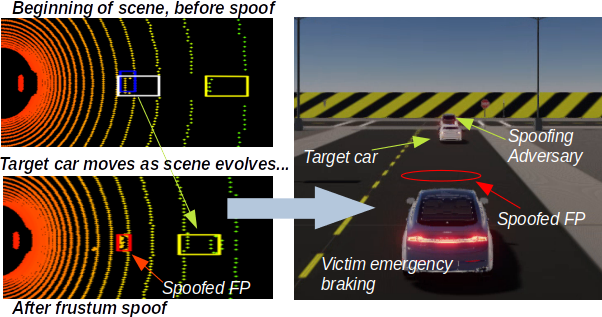}
    \vspace{-8pt}
    \caption{Frustum attack achieved on AV running Baidu Apollo software using perception data from LGSVL simulator with setup matching the physical experiment from Section~\ref{sec:3-attack-model-lidar-spoofing}. Target vehicle (white) detected throughout by image-detection pipeline. Target vehicle initially detected by LiDAR perception (blue) and followed by Apollo. When spoofing happens, Apollo detects spoofed points as new object (red); thus, unnecessarily engages emergency brakes and stops mid-lane.}
    \vspace{-8pt}
    \label{fig:baidu-apollo-1}
\end{figure}

The scene is set consistent with the physical experiment (Section~\ref{sec:3-attack-model-lidar-spoofing}) and following Fig.~\ref{fig:spoof-scenarios-2} (i.e.,~\textbf{S2}): a target car is between the victim and the spoofing adversary. Fig.~\ref{fig:baidu-apollo-1} shows snapshots of two LiDAR captures with detected objects when running Apollo perception (left) and the control outcome observed when replayed through LGSVL (right). Initially, Apollo detects the target car, as expected. The spoofing adversary is not detected due to strong occlusion, also as expected. Part-way through the sequence, the adversary launches the frustum attack, and Apollo detects the spoofed points as a nearby object which triggers emergency braking, unnecessarily stopping (and thus endangering the victim vehicle). Full video of the playback sequence is available~at~\cite{TheAttack}.


\paragraph{Remarks.}
Attacks on perception must propagate into adversarial tracks to impact AVs. We have shown that frustum attacks can be exercised longitudinally to have high-impact at the tracking, decision, and control levels. An attacker can use mere seconds of real-time to create false scenarios of predicted collision or accelerate the flow of traffic. The frustum attack allows for both starting attacks at longer range and attacking in front-near. We show that this can be of great benefit for the attacker because it can create a diverse set of attacker-specified, high-confidence maneuvers.
\vspace{-2pt}
\section{Discussion and Future Work}
\label{sec:8-discussion}


\subsection{Limitations}
\paragraph{Datasets.} We use KITTI and LGSVL to evaluate the considered LiDAR spoofing attacks. Although we generated over 75 million attack scenarios, KITTI is a small dataset and may not be fully representative of day-to-day AV driving. We use the LGSVL simulator to study Baidu Apollo, and future work will leverage high-fidelity simulators and additional open-source datasets to perform studies on frustum attack generalization.

\paragraph{Apollo Evasive Maneuvers.} 
Besides our testing of the frustum attack in front of the target vehicle on Apollo (Sec.~\ref{sec:7-longitudinal-baidu-apollo}), we intended to test the frustum attack in the shadow region (i.e.,~behind target) since the shadow is more vulnerable to attack, as identified in Sec.~\ref{sec:6-frustum-attack-results}. However, this was not attainable with Apollo's capabilities. We observed that Apollo has minimal ability to execute any evasive maneuvers, even when we aimed a target car heading straight for Apollo using ground-truth perception data. This is consistent with findings from 
evaluations of limitations of Apollo capabilities~\cite{Piazzoni2021ViSTA:Vehicles}.


\paragraph{Optical Engineering and Dynamic Spoofing.} The frustum attack is logistically possible in the scenarios described in Sec.~\ref{sec:3-attack-model-lidar-spoofing}. While these 
commonly occur in everyday driving, only three are technically feasible with today's technology, and only two have thus been demonstrated - those also used static spoofer and static victim. The current experiments have not shown attack feasibility when there are relative distance and angle changes between the spoofer and the victim. In all five targeted attack scenarios in Fig.~\ref{fig:spoof-scenarios}, the victim should be moving to cause serious attack consequences. This would require the spoofing device to dynamically track and aim at the victim, and this engineering feat has not yet been fully demonstrated, with some recent progress in this direction~\cite{Cao2021InvisibleAttacks}.

\vspace{-2pt}
\subsection{Future Work}
\paragraph{Shadow Vulnerability.} The shadow region, a subset of the frustum behind the target object, is an important element for the frustum attack success. Future work will explore in detail the low-level behavior of the perception DNNs to illuminate why the shadow is so vulnerable, including possibilities of overfitting or intrinsic vulnerability of free space. 

\vspace{-2pt}
\paragraph{Defenses.} We evaluated state-of-the-art defenses against LiDAR spoofing and found none are suitable to protect against the frustum attack. Future works will propose defenses capable of defending against the frustum~attack.

\vspace{-2pt}
\paragraph{Generalization.} The high degree of frustum attack success and the large number of evaluations performed suggests that a single choice of attack parameters can generalize across perception algorithms. However, this belief has only been tested implicitly using consistent attack parameters in the large-scale study. Future works will explicitly consider the success of transferring specific attack traces between algorithms.

\vspace{-4pt}
\section{Conclusion}
\label{sec:9-conclusion}
\vspace{-4pt}

In this work, we exposed the vulnerability of LiDAR-only perception and camera-LiDAR fusion to the frustum attack: small-scale (i.e., tens of points) LiDAR spoofing in-view of existing, valid objects. We evaluated the frustum attack on three distinct LiDAR-only architectures and five models within three different architectures of camera-LiDAR fusion, including fusion at the semantic, feature, and tracking levels. Within each class, we used single-sensor and sensor fusion algorithms from top-performers on popular datasets (\cite{Huang2020Epnet:Detection, Ku2018, Wang2019c, Lang2019a, Shi2019, Yang2018a}), established benchmarks (\cite{Ku2018, Qi2018b, Shi2019, Lang2019a}), and algorithms representative of leading end-to-end, full-stack industry pipelines (\cite{2021BaiduApollo, Shi2019, Lang2019a}). We demonstrated a singular attack model capable of compromising each class perception in AVs. The attack model is black-box; furthermore, it does not require any knowledge of the perception algorithm. Such broad success with a black-box attack model illuminates a systematic vulnerability across both LiDAR-only and camera-LiDAR perception algorithms.

\vspace{-2pt}
\subsection*{Acknowledgements}
This work is sponsored in part by the ONR under agreement N00014-20-1-2745, AFOSR award number FA9550-19-1-0169, and NSF CNS-1652544 and CNS-2112562 awards. 

\setlength\bibitemsep{3.2pt}
\bibliographystyle{ieeetr}
\bibliography{references_pruned,references-links-only}

\begin{thebibliography}{10}

\bibitem{Hawkins2018}
A.~Hawkins, ``{Waymo’s autonomous cars have driven 8 million miles on public
  roads}.''
  \url{https://www.theverge.com/2018/7/20/17595968/waymo-self-driving-cars-8-million-miles-testing},
  2018.

\bibitem{2021TheTechnologies}
``{The Evolution of Automated Safety Technologies}.''
  \url{https://www.nhtsa.gov/technology-innovation/automated-vehicles-safety},
  2021.

\bibitem{Hecht2018}
J.~Hecht, ``{Lidar for self-driving cars},'' {\em Optics and Photonics News},
  vol.~29, no.~1, pp.~26--33, 2018.

\bibitem{2017GMDeveloper}
``{GM Advances Self-Driving Vehicle Deployment With Acquisition of LIDAR
  Developer}.''
  \url{https://media.gm.com/media/us/en/gm/news.detail.html/content/Pages/news/us/en/2017/oct/1009-lidar1.html},
  2017.

\bibitem{2021NVIDIADRIVE}
``{NVIDIA DRIVE}.'' \url{https://developer.nvidia.com/drive}.

\bibitem{2021BaiduApollo}
``{Baidu Apollo}.'' \url{apollo.auto}.

\bibitem{Schoettle2015}
B.~Schoettle and M.~Sivak, ``{A preliminary analysis of real-world crashes
  involving self-driving vehicles},'' {\em Univ. of Michigan Transportation
  Research Institute}, 2015.

\bibitem{Kohli2019}
P.~Kohli and A.~Chadha, ``{Enabling pedestrian safety using computer vision
  techniques: A case study of the 2018 uber inc. self-driving car crash},'' in
  {\em Future of Information and Communication Conf.}, pp.~261--279, 2019.

\bibitem{Eykholt2018b}
K.~Eykholt, I.~Evtimov, E.~Fernandes, B.~Li, A.~Rahmati, C.~Xiao, A.~Prakash,
  T.~Kohno, and D.~Song, ``{Robust Physical-World Attacks on Deep Learning
  Visual Classification},'' in {\em Proceedings of the IEEE Computer Society
  Conference on Computer Vision and Pattern Recognition}, pp.~1625--1634, 2018.

\bibitem{Cao2019d}
Y.~Cao, C.~Xiao, B.~Cyr, Y.~Zhou, W.~Park, S.~Rampazzi, Q.~A. Chen, K.~Fu, and
  Z.~M. Mao, ``{Adversarial sensor attack on lidar-based perception in
  autonomous driving},'' in {\em Proc. of the 2019 ACM SIGSAC Conf. on Computer
  and Communications Security}, pp.~2267--2281, 2019.

\bibitem{Sun2020h}
J.~Sun, Y.~Cao, Q.~A. Chen, and Z.~Morley~Mao, ``{Towards robust LiDAR-based
  perception in autonomous driving: General black-box adversarial sensor attack
  and countermeasures},'' in {\em Proceedings of the 29th USENIX Security
  Symposium}, pp.~877--894, 2020.

\bibitem{Tu2021b}
J.~Tu, H.~Li, X.~Yan, M.~Ren, Y.~Chen, M.~Liang, E.~Bitar, E.~Yumer, and
  R.~Urtasun, ``{Exploring Adversarial Robustness of Multi-Sensor Perception
  Systems in Self Driving},'' {\em arXiv preprint arXiv:2101.06784}, 2021.

\bibitem{Abdelfattah2021}
M.~Abdelfattah, K.~Yuan, Z.~J. Wang, and R.~Ward, ``{Adversarial Attacks on
  Camera-LiDAR Models for 3D Car Detection},'' {\em arXiv preprint
  arXiv:2103.09448}, 2021.

\bibitem{Petit2015c}
J.~Petit, B.~Stottelaar, M.~Feiri, and F.~Kargl, ``{Remote Attacks on Automated
  Vehicles Sensors: Experiments on Camera and LiDAR},'' {\em Blackhat.com},
  vol.~11, pp.~1--13, 2015.

\bibitem{Shin2017d}
H.~Shin, D.~Kim, Y.~Kwon, and Y.~Kim, ``{Illusion and dazzle: Adversarial
  optical channel exploits against lidars for automotive applications},'' in
  {\em Lecture Notes in Computer Science}, vol.~10529 LNCS, pp.~445--467, 2017.

\bibitem{Cao2021AutomatedTargets}
Y.~Cao, J.~Ma, K.~Fu, R.~Sara, and M.~Mao, ``{Automated Tracking System For
  LiDAR Spoofing Attacks On Moving Targets},'' 2021.

\bibitem{Cao2021InvisibleAttacks}
Y.~Cao, N.~Wang, C.~Xiao, D.~Yang, J.~Fang, R.~Yang, Q.~A. Chen, M.~Liu, and
  B.~Li, ``{Invisible for both Camera and LiDAR: Security of Multi-Sensor
  Fusion based Perception in Autonomous Driving Under Physical-World
  Attacks},'' in {\em 2021 IEEE Symposium on Security and Privacy (SP)},
  pp.~176--194, 2021.

\bibitem{Liu2021}
J.~Liu and J.~Park, ``{" Seeing is not Always Believing": Detecting Perception
  Error Attacks Against Autonomous Vehicles},'' {\em IEEE Transactions on
  Dependable and Secure Computing}, 2021.

\bibitem{Pajic2014b}
M.~Pajic, J.~Weimer, N.~Bezzo, P.~Tabuada, O.~Sokolsky, I.~Lee, and G.~J.
  Pappas, ``{Robustness of attack-resilient state estimators},'' in {\em 2014
  ACM/IEEE Int. Conf. on Cyber-Physical Systems (ICCPS)}, pp.~163--174, 2014.

\bibitem{Hau2020Shadow-Catcher:Sensing}
Z.~Hau, S.~Demetriou, L.~Mu{\~{n}}oz-Gonz{\'{a}}lez, and E.~C. Lupu,
  ``{Shadow-Catcher: Looking Into Shadows to Detect Ghost Objects in Autonomous
  Vehicle 3D Sensing},'' {\em arXiv preprint arXiv:2008.12008}, 2020.

\bibitem{Rong2020LgsvlDriving}
G.~Rong, B.~H. Shin, H.~Tabatabaee, Q.~Lu, S.~Lemke, M.~Mo{\v{z}}eiko,
  E.~Boise, G.~Uhm, M.~Gerow, and S.~Mehta, ``{Lgsvl simulator: A high fidelity
  simulator for autonomous driving},'' in {\em 2020 IEEE 23rd Int. Conf. on
  Intelligent Transportation Systems (ITSC)}, pp.~1--6, 2020.

\bibitem{Geiger2013b}
A.~Geiger, P.~Lenz, C.~Stiller, and R.~Urtasun, ``{Vision meets robotics: The
  KITTI dataset},'' {\em Int. Journal of Robotics Research}, vol.~32, no.~11,
  pp.~1231--1237, 2013.

\bibitem{Kato2018AutowareSystems}
S.~Kato, S.~Tokunaga, Y.~Maruyama, S.~Maeda, M.~Hirabayashi, Y.~Kitsukawa,
  A.~Monrroy, T.~Ando, Y.~Fujii, and T.~Azumi, ``{Autoware on board: Enabling
  autonomous vehicles with embedded systems},'' in {\em 2018 ACM/IEEE 9th
  International Conference on Cyber-Physical Systems (ICCPS)}, pp.~287--296,
  IEEE, 2018.

\bibitem{Kutila2016}
M.~Kutila, P.~Pyyk{\"{o}}nen, W.~Ritter, O.~Sawade, and B.~Sch{\"{a}}ufele,
  ``{Automotive LIDAR sensor development scenarios for harsh weather
  conditions},'' in {\em 19th IEEE ITSC}, pp.~265--270, 2016.

\bibitem{Liang2018a}
M.~Liang, B.~Yang, S.~Wang, and R.~Urtasun, ``{Deep Continuous Fusion for
  Multi-sensor 3D Object Detection},'' in {\em Lecture Notes in Computer
  Science}, vol.~11220 LNCS, pp.~663--678, 2018.

\bibitem{Zhou2018a}
Y.~Zhou and O.~Tuzel, ``{VoxelNet: End-to-End Learning for Point Cloud Based 3D
  Object Detection},'' in {\em Proc. of the IEEE Computer Society Conference on
  Computer Vision and Pattern Recognition}, pp.~4490--4499, 2018.

\bibitem{Lang2019a}
A.~H. Lang, S.~Vora, H.~Caesar, L.~Zhou, J.~Yang, and O.~Beijbom,
  ``{Pointpillars: Fast encoders for object detection from point clouds},'' in
  {\em Proc. of the IEEE Computer Society Conf. on Computer Vision and Pattern
  Recognition}, vol.~2019-June, pp.~12689--12697, 2019.

\bibitem{Shi2019}
S.~Shi, X.~Wang, and H.~Li, ``{Pointrcnn: 3d object proposal generation and
  detection from point cloud},'' in {\em Proceedings of the IEEE/CVF Conference
  on Computer Vision and Pattern Recognition}, pp.~770--779, 2019.

\bibitem{Qi2018b}
C.~R. Qi, W.~Liu, C.~Wu, H.~Su, and L.~J. Guibas, ``{Frustum PointNets for 3D
  Object Detection from RGB-D Data},'' in {\em Proceedings of the IEEE Computer
  Society Conference on Computer Vision and Pattern Recognition}, pp.~918--927,
  2018.

\bibitem{Wang2019c}
Z.~Wang and K.~Jia, ``{Frustum ConvNet: Sliding Frustums to Aggregate Local
  Point-Wise Features for Amodal},'' {\em IEEE International Conference on
  Intelligent Robots and Systems}, pp.~1742--1749, 2019.

\bibitem{Ku2018}
J.~Ku, M.~Mozifian, J.~Lee, A.~Harakeh, and S.~L. Waslander, ``{Joint 3D
  Proposal Generation and Object Detection from View Aggregation},'' in {\em
  IEEE Int. Conference on Intelligent Robots and Systems}, pp.~5750--5757,
  2018.

\bibitem{Huang2020Epnet:Detection}
T.~Huang, Z.~Liu, X.~Chen, and X.~Bai, ``{Epnet: Enhancing point features with
  image semantics for 3d object detection},'' in {\em European Conference on
  Computer Vision}, pp.~35--52, Springer, 2020.

\bibitem{Yang2018a}
B.~Yang, W.~Luo, and R.~Urtasun, ``{PIXOR: Real-time 3D Object Detection from
  Point Clouds},'' in {\em Proceedings of the IEEE Computer Society Conference
  on Computer Vision and Pattern Recognition}, pp.~7652--7660, 2018.

\bibitem{Papernot2017b}
N.~Papernot, P.~McDaniel, I.~Goodfellow, S.~Jha, Z.~B. Celik, and A.~Swami,
  ``{Practical black-box attacks against machine learning},'' in {\em ASIA CCS
  2017 - Proceedings of the 2017 ACM Asia Conference on Computer and
  Communications Security}, pp.~506--519, 2017.

\bibitem{Boloor2020b}
A.~Boloor, K.~Garimella, X.~He, C.~Gill, Y.~Vorobeychik, and X.~Zhang,
  ``{Attacking vision-based perception in end-to-end autonomous driving
  models},'' {\em Journal of Systems Architecture}, vol.~110, p.~101766, 2019.

\bibitem{Ivanov2014b}
R.~Ivanov, M.~Pajic, and I.~Lee, ``{Attack-resilient sensor fusion},'' in {\em
  2014 Design, Automation {\&} Test in Europe Conference {\&} Exhibition
  (DATE)}, pp.~1--6, IEEE, 2014.

\bibitem{TheAttack}
``{The Frustum Attack}.''
  \url{https://cpsl.pratt.duke.edu/research/frustum-attack}.

\bibitem{Blackman1986}
S.~S. Blackman, ``{Multiple-target tracking with radar applications},'' {\em
  Dedham}, 1986.

\bibitem{CommaAI}
``{Comma AI}.'' \url{https://comma.ai/}.

\bibitem{Li2003}
X.~R. Li and V.~P. Jilkov, ``{Survey of maneuvering target tracking. Part I.
  Dynamic models},'' {\em IEEE Transactions on Aerospace and Electronic
  Systems}, vol.~39, no.~4, pp.~1333--1364, 2003.

\bibitem{Jia2020a}
Y.~Jia, Y.~Lu, J.~Shen, Q.~A. Chen, Z.~Zhong, and T.~Wei, ``{Fooling Detection
  Alone is Not Enough: First Adversarial Attack against Multiple Object
  Tracking},'' in {\em Int. Conference on Learning Representations (ICLR)},
  2020.

\bibitem{Piazzoni2021ViSTA:Vehicles}
A.~Piazzoni, J.~Cherian, M.~Azhar, J.~Y. Yap, J.~L.~W. Shung, and R.~Vijay,
  ``{ViSTA: a Framework for Virtual Scenario-based Testing of Autonomous
  Vehicles},'' {\em arXiv preprint arXiv:2109.02529}, 2021.

\end{thebibliography}

\appendix
\section{
Existing Attacks and Defenses}
\label{sec:appendix-reproduce}


\subsection{Naive Spoofing 
on LiDAR-only Perception} 
\label{sec:appendix-reproduce-attacks}

We implement the naive spoofing of~\cite{Cao2019d, Sun2020h} with the attack model described in Section~\ref{sec:3-attack-model-lidar-spoofing}. We follow~\cite{Sun2020h} to evaluate the attack success; i.e., we selected 5 attack traces using $\{10,\,20,\,...200\}$ points per trace over multiple trials for 100 attack evaluations, and placed the spoofed points in front-near positions, $5-8~m$ from the victim vehicle. We extend the evaluation to outside of the front-near when evaluating defenses. Attack Success Rate (ASR) for FP outcomes is defined as the fraction of times a fictitious object is detected over the number of targeted attempts (e.g., number of spoof point clusters), as there could be more than one FP per frame. Similarly, we define the FN ASR as the fraction of times an object is missed over the number of attempts. 

\begin{figure}[!t]
    \centering
    \includegraphics[width=0.76\linewidth]{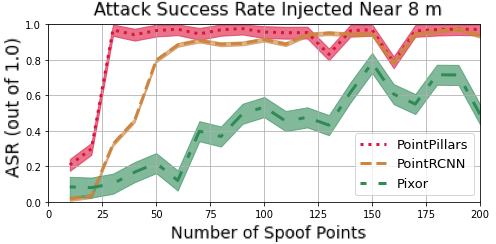} 
    \vspace{-8pt}
    \caption{\textit{Naive spoofing attacks against LiDAR-only perception:} Reproduced naive black-box spoofing attacks from~\cite{Sun2020h} applied to LiDAR-only perception, one method from each of the three LiDAR-only architecture categories from Table~\ref{tab:perception-algs}.}
    \label{fig:naive-spoof-8m}
    \vspace{-6pt}
\end{figure}

Our results, summarized in Fig.~\ref{fig:naive-spoof-8m}, confirm the success of the naive black-box spoofing attacks. The ASR is consistently high when 60 or more spoof points are used for all LiDAR-only algorithms showing that each of the 3 architectures are deeply vulnerable to spoof injections in front-near~positions.

\subsection{Existing Defenses against LiDAR Spoofing} 
\label{sec:appendix-reproduce-defenses}

\begin{figure}[!t]
    \centering
    \includegraphics[width=0.76\linewidth]{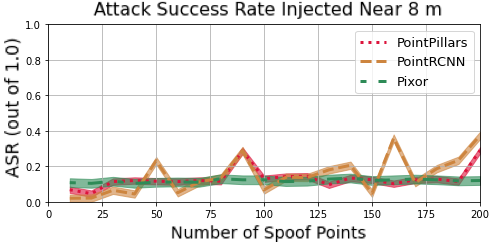} 
    \vspace{-8pt}
    \caption{\textit{Naive spoofing attacks on LiDAR-only perception with CARLO}: CARLO 
    guards LiDAR-only perception against naive black-box spoofing attacks \textbf{only} in front-near positions.}
    \label{fig:naive-carlo-8m}
    \vspace{-6pt}
\end{figure}

\begin{figure}[!t]
    \centering
    \includegraphics[width=0.76\linewidth]{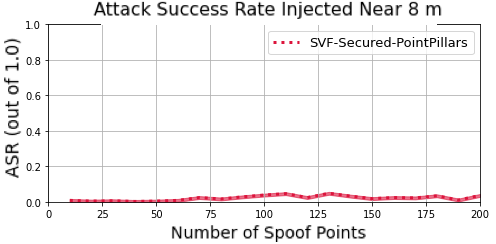} 
    \vspace{-8pt}
    \caption{\textit{Naive spoofing attacks on LiDAR-only perception with SVF:} 
    SVF guards LiDAR-only perception against naive black-box spoofing in front-near positions. SVF requires rearchitecting the perception model which is not feasible for every algorithm. We test SVF-modified PointPillars.}
    \label{fig:svf-naive}
    \vspace{-6pt}
\end{figure}

We reproduce the state-of-the-art defenses against LiDAR spoofing; we showed that both CARLO and SVF dramatically reduce the ASR in front-near positions, while ShadowCatcher has challenges defending naive attacks (see 
Figs.~\ref{fig:naive-carlo-8m} and~\ref{fig:svf-naive}). Note that SVF requires model-level changes and expensive retraining which are not possible with all the tested perception algorithms. Thus, we rearchitect and retrain PointPillars with SVF, following the approach from~\cite{Sun2020h}.

\begin{figure}[!t]
    \centering
    \includegraphics[width=0.76\linewidth]{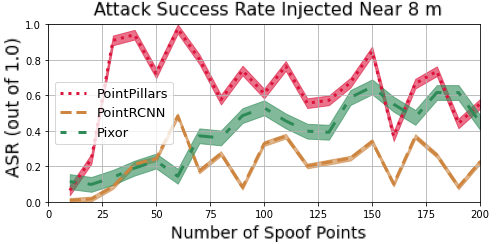} 
    \vspace{-8pt}
    \caption{\textit{Naive spoofing attacks on LiDAR-only perception with ShadowCatcher:} 
    ShadowCatcher has limited ability to guard against LiDAR spoofing attacks presented in~\cite{Sun2020h} due to difficulty handling noisy shadow estimation.}
    \label{fig:gbsc-naive}
    \vspace{-8pt}
\end{figure}

While the ShadowCatcher defense is impressively simple, we do not expect to obtain the high accuracy reported in~\cite{Hau2020Shadow-Catcher:Sensing}. The reason is that the original work made several assumptions that are unrealistic, including tuning parameters on the test set, using ground truth bounding boxes instead of outputs of a perception algorithm, which significantly alters the shadow-region estimation noise, and only testing on 200 scenes with only three selections of 200 spoofed points. Still, we reproduced the original ShadowCatcher defense and obtained not as strong detection results without these assumptions 
(Fig.~\ref{fig:gbsc-naive}).

\section{CARLO Vulnerabilities}
\label{sec:appendix-defenses-vulnerable}
\vspace{-4pt}

We described in Section~\ref{sec:5-naive-spoofing-defenses} that CARLO 
introduces vulnerability to FP attacks outside front-near and FN invalidation attacks. Here, we provide additional quantitative analysis. 

\begin{figure}[!t]
    \centering
    \begin{subfigure}[c]{0.48\linewidth} 
      \centering
      \includegraphics[width=1\textwidth]{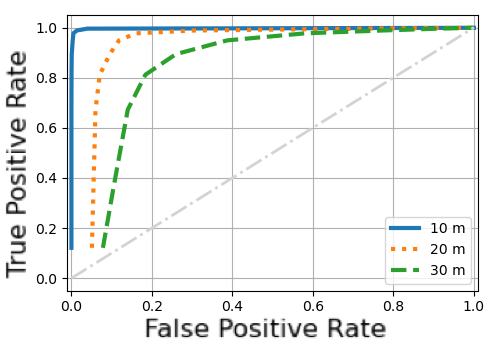} 
      \vspace{-16pt}
    \end{subfigure}
    \begin{subfigure}[c]{0.48\linewidth}  
      \centering
      \includegraphics[width=1\textwidth]{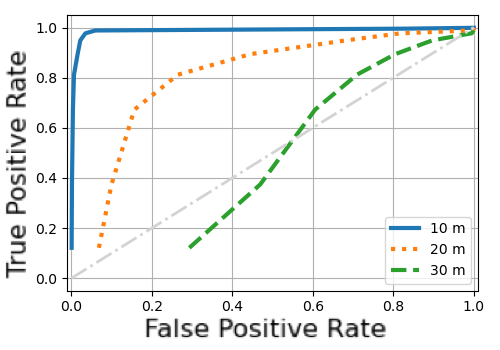} 
      \vspace{-16pt}
    \end{subfigure}
    \vspace{-6pt}
    \caption{ROC for FP attack on CARLO using PointPillars as perception with (left) 60 and (right) 200 spoofed points. 
    }
    \vspace{-9pt}
    \label{fig:naive-carlo-fp-roc}
\end{figure}

To evaluate CARLO outside front-near, we collect receiver operating characteristics (ROC) on CARLO's ability to distinguish between valid and spoof objects placed at different ranges from the victim AV; the results are presented in Fig.~\ref{fig:naive-carlo-fp-roc}. As in~\cite{Sun2020h}, we use PointPillars for this test; yet, CARLO is model-agnostic and these results generalize across other algorithms. We observe that, as range of the spoofed objects increases, CARLO's classification performance deteriorates (i.e., the defense breaks) and the ROC curve moves towards the center; e.g., in the case of 200 spoof points at 30~m (green), CARLO is similar to the random-guessing classifier.

\begin{figure}[!t]
    \centering
    \includegraphics[width=0.62\linewidth]{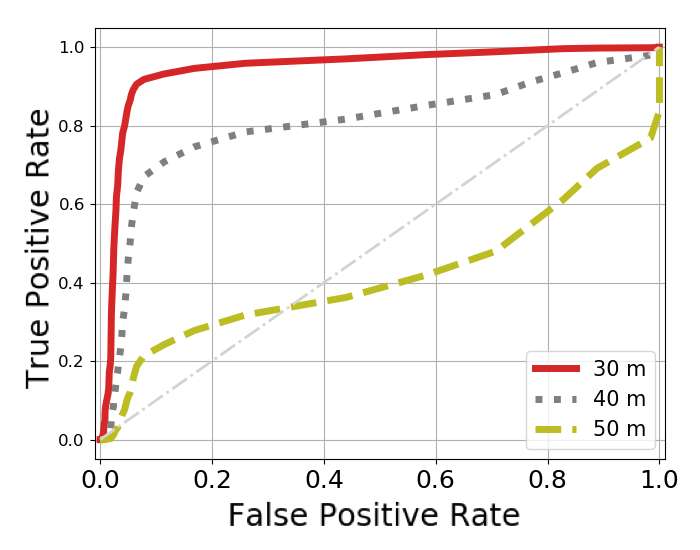}
    \vspace{-12pt}
    \caption{The CARLO defense is vulnerable to 
    FN attacks. Classification of valid objects 
    significantly degrades when 
    randomly spoofing 200 points behind valid objects, 
    as well as when 
    range to target object increases.} 
    \label{fig:naive-carlo-fn-roc}
    \vspace{-8pt}
\end{figure}

Similarly, to test the FN invalidation attack, we collect ROC curves where each curve represents the range to the targeted valid object in Fig.~\ref{fig:naive-carlo-fn-roc}. For each object, we record the range to that object, add 200 spoof points  
in a random pattern behind~it, run CARLO on the detected result, 
and~check~if~it
invalidated the true object. We find that CARLO's performance against the invalidation attack deteriorates when true objects are at an increased range from the victim, likely due to the decreased density of LiDAR points when objects are farther away.

\vspace{-4pt}
\section{
Spoofing Scenarios for Frustum Attacks}
\label{sec:appendix-spoofing-feasibility}
\vspace{-4pt}

\begin{figure}[!t]
    \centering
    \begin{subfigure}[t]{.98\linewidth}
      \centering
      \includegraphics[width=0.7\textwidth]{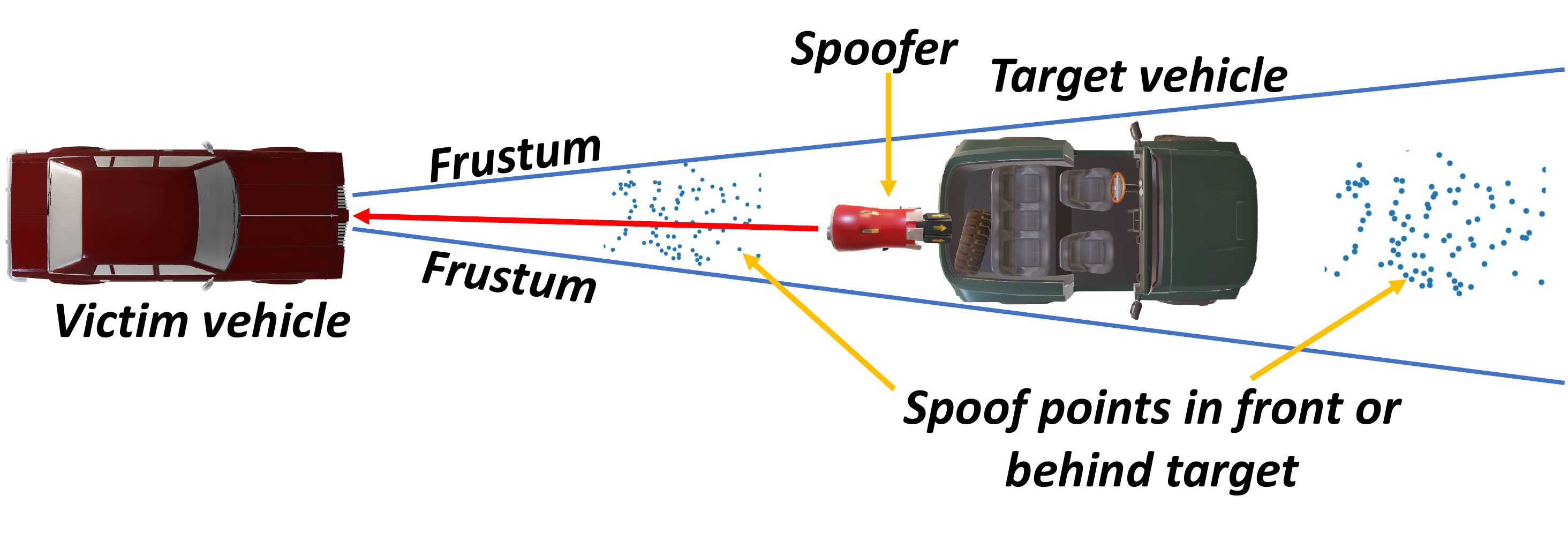} 
      \vspace{-6pt}
      \caption{Spoofer placed on target vehicle and points spoofed in LOS.} 
      \label{fig:spoof-scenarios-1}
    \end{subfigure}
    \begin{subfigure}[t]{.98\linewidth}
      \centering
      \includegraphics[width=0.7\textwidth]{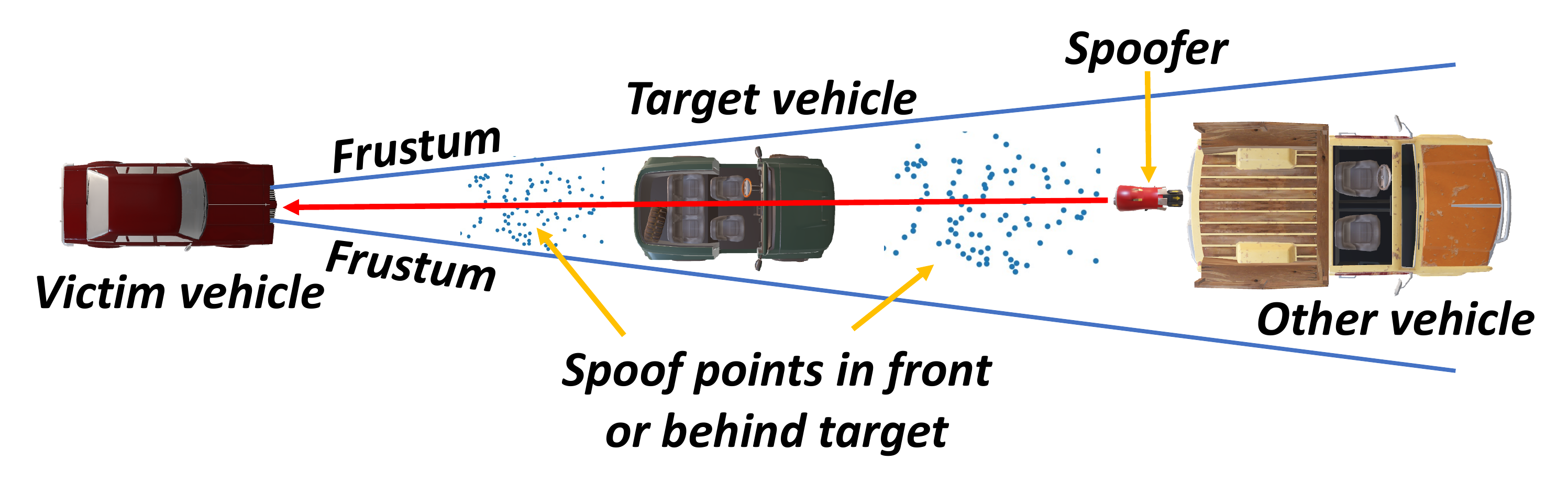}   
      \vspace{-6pt}
      \caption{Spoofer placed on larger vehicle in front of target, pointing in LOS.}
      \label{fig:spoof-scenarios-2}
    \end{subfigure}
       \begin{subfigure}[t]{.98\linewidth}
       \vspace{-12pt}
      \centering
      \includegraphics[width=0.64\textwidth]{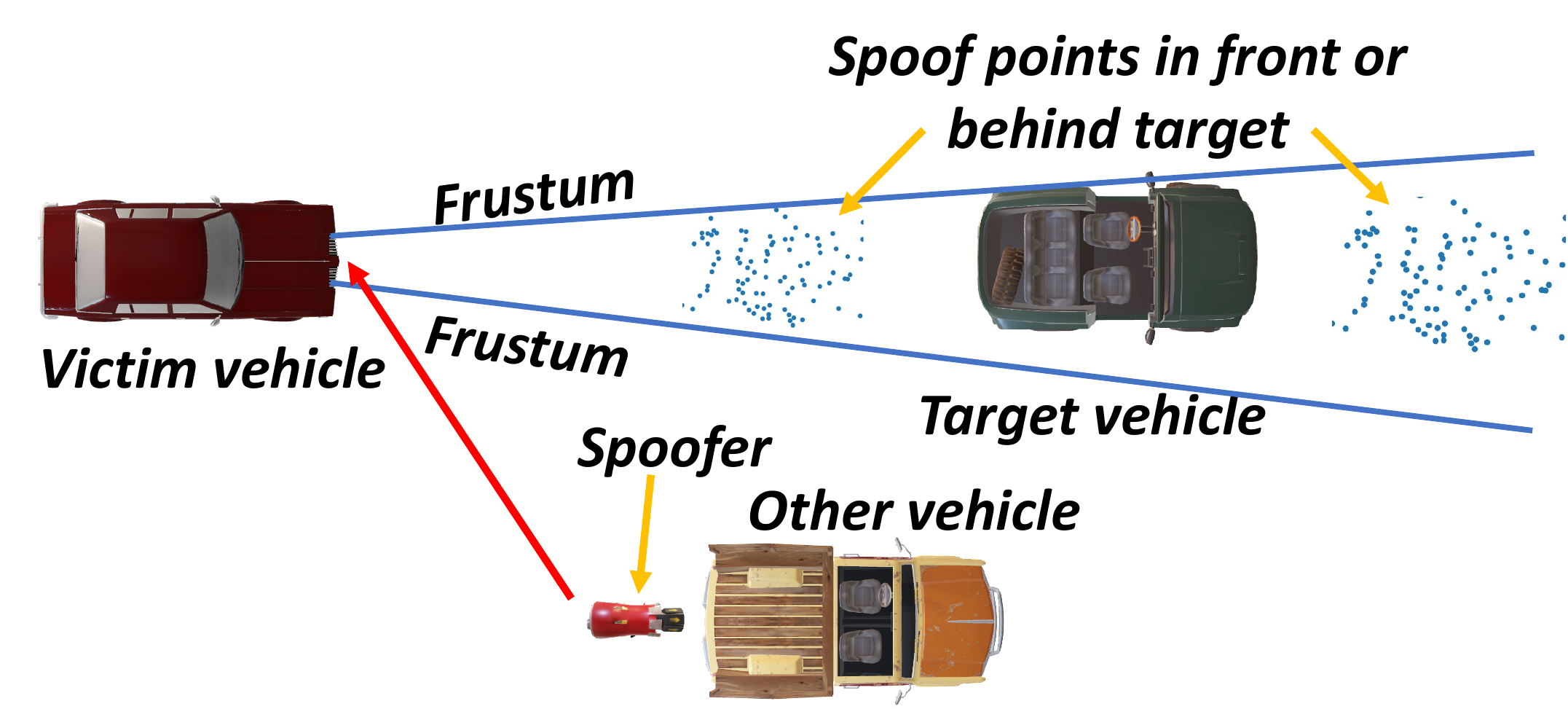}  
      \vspace{-6pt}
      \caption{Spoofer placed on other vehicle in nearby lane, pointing out~of LOS.}
      \label{fig:spoof-scenarios-3}
    \end{subfigure}
    \begin{subfigure}[t]{.98\linewidth}
      \vspace{-12pt}
      \centering
      \includegraphics[width=0.68\textwidth]{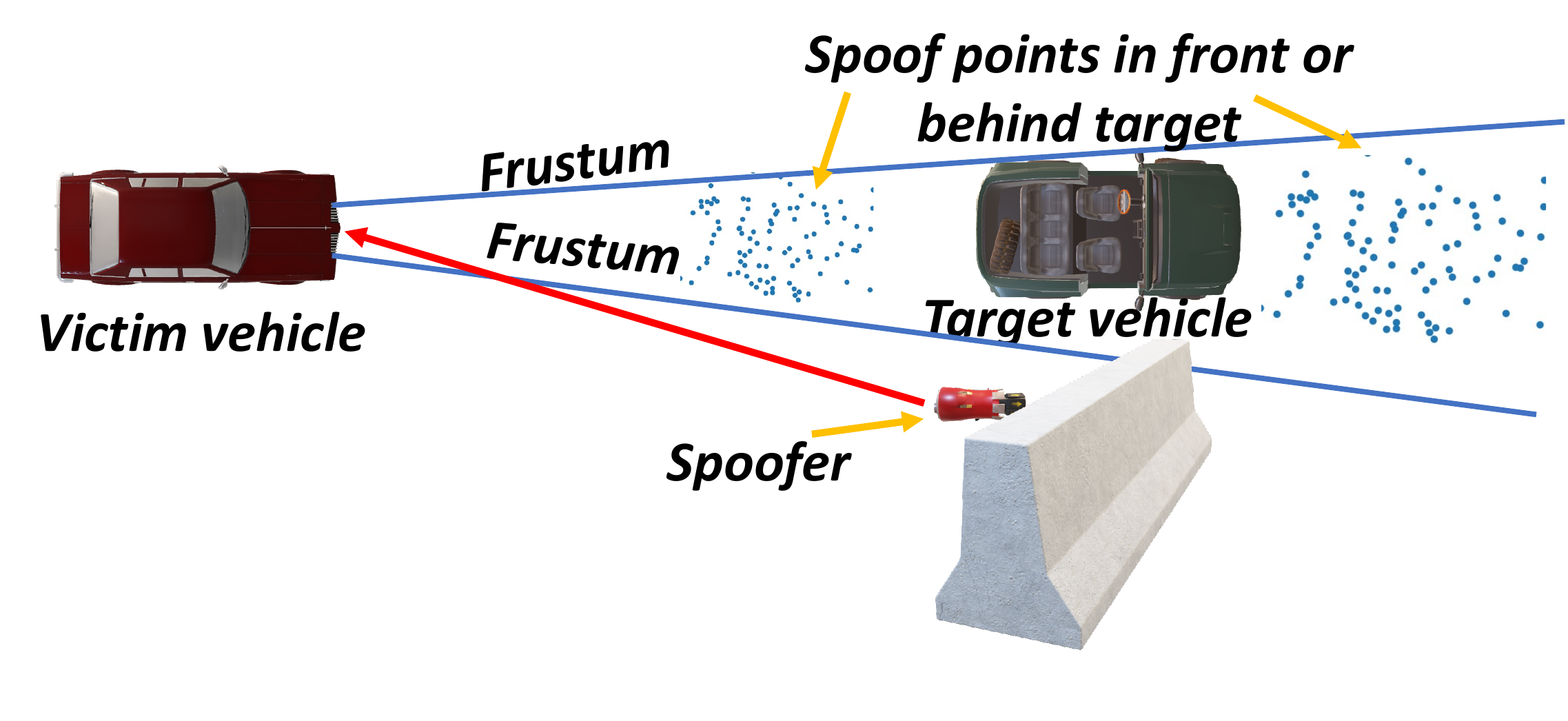}    
      \vspace{-14pt}
      \caption{Spoofer placed in environment e.g.,~on roadside obstacle, 
      in LOS (e.g.,~on bridge) or out of LOS (e.g.,~on roadside)}
      \label{fig:spoof-scenarios-4}
    \end{subfigure}
    \vspace{-8pt}
    \caption{The frustum attack 
    in everyday driving scenarios. Adversaries can 
    place a spoofer on target car, on other cars, or on roadside obstacles, 
    placing points anywhere along line-of-sight (LOS) (e.g.,~in front or behind the target~car).}
    \label{fig:spoof-scenarios}
    \vspace{-10pt}
\end{figure}

We illustrate common 
situations where vehicle configurations enable frustum attack spoofing (Fig.~\ref{fig:spoof-scenarios}); the specific scenarios are described in Sec.~\ref{sec:3-attack-model-lidar-spoofing}. We execute the physical experiment in Section~\ref{sec:3-attack-model-lidar-spoofing} corresponding to the scenario in Fig.~\ref{fig:spoof-scenarios-2} and similar to the scenario in Fig.~\ref{fig:spoof-scenarios-1}. We also perform the longitudinal case study in Sec.~\ref{sec:7-longitudinal-baidu-apollo} in accordance with the scenario from~Fig.~\ref{fig:spoof-scenarios-2}. Anticipated advances in optical technology and tracking will soon enable spoofing points outside of line-of-sight 
as demonstrated in e.g., \cite{Cao2021AutomatedTargets}.

\vspace{-5pt}
\section{Impact of Spoof Point Placement}
\label{sec:appendix-spoof-placement}
\vspace{-4pt}

Prior 
works required spoofed points to be placed in patterns of occluded vehicles~\cite{Sun2020h}. Here, we show that spoofing in a normally-distributed pattern for the frustum attack can have success nearly matching using occluded traces. 

\begin{figure}[!t]
    \begin{subfigure}[t]{0.48\linewidth}
      \centering
      \includegraphics[width=\textwidth]{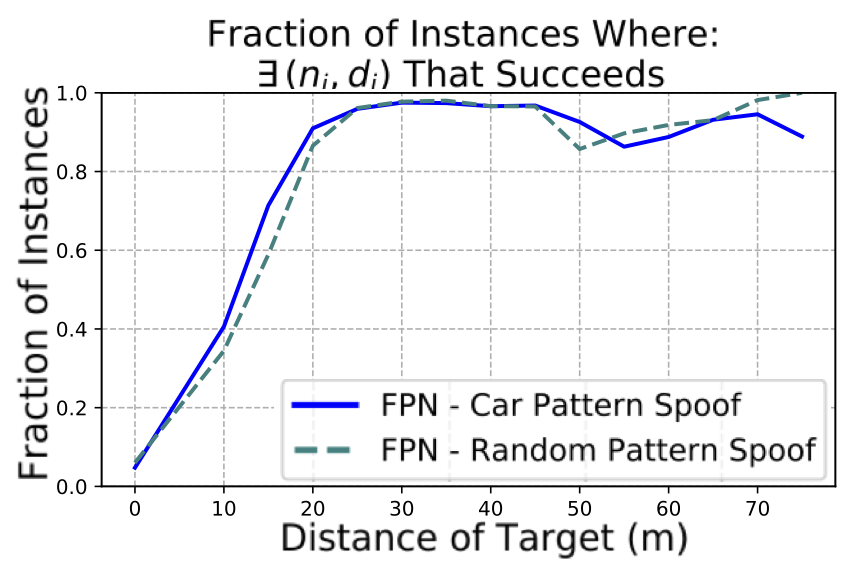}  
      \label{fig:comparison-spoof-pattern-1}
    \end{subfigure}
    \begin{subfigure}[t]{0.48\linewidth}
      \centering
      \includegraphics[width=\linewidth]{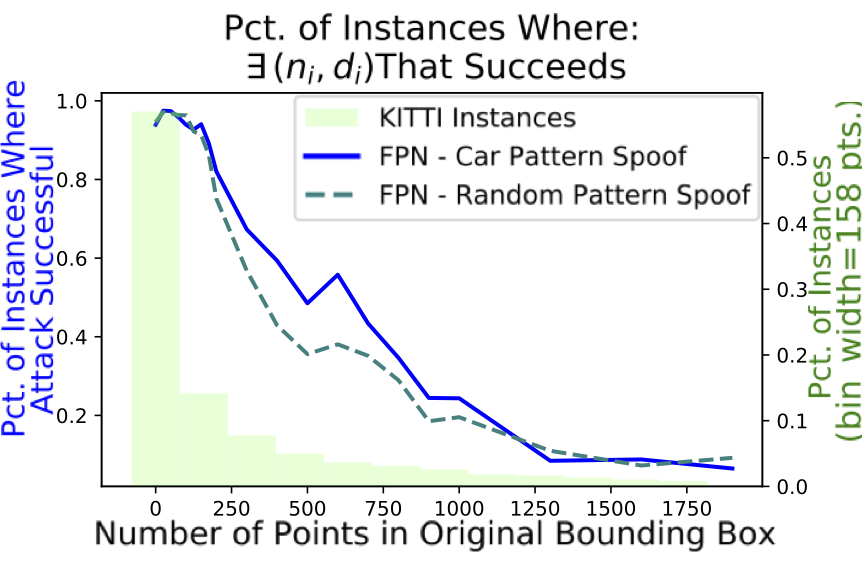}    
      \label{fig:comparison-spoof-pattern-2}
    \end{subfigure}
    \vspace{-22pt}
    \caption{Spoofing in a Gaussian random pattern (Table~\ref{tab:point-params}) achieves performance on-par with using an occluded car pattern; we test spoofing patterns~on~FPN 
    and show dependence of attackability on (left) range to target and (right) number of points in target object bounding box with histogram showing frequency (\%) of occurrence of such objects in KITTI.}
    \label{fig:spoof-point-pattern-comparison}
    \vspace{-6pt}
\end{figure}

In Fig.~\ref{fig:spoof-point-pattern-comparison}, we compare the attackability of FPN perception when using the two spoof point generation methods (similar results are also obtained for the other aforementioned perception algorithms). We first provide a comparison of the fraction of instances where an attack succeeds using both methods, as function of the distance to the target (Fig.~\ref{fig:spoof-point-pattern-comparison}-left)
As can be observed, it is difficult to distinguish between the performance of the two methods (random points vs. car-pattern). 

We further provide an analysis of the attackability as a function of the number of points in the target objects' bounding box (Fig.~\ref{fig:spoof-point-pattern-comparison}-right)
We define attackability as the ability to find an attack that succeeds within the attacker-specified capabilities. 
Both methods perform similarly, 
with a small benefit of car patterned injections at medium range.
%
This improve the feasibility of LiDAR-based spoofing, as a normally distributed pattern does not require unrealistically careful placement of spoof points, and 
is robust to small displacements. 


\vspace{-4pt}
\section{Frustum Attack By Range to Target}
\label{sec:appendix-frustum-attack-all-algorithms}
\vspace{-4pt}

\begin{figure*}[t!]
    \centering
    \includegraphics[width=0.88\textwidth]{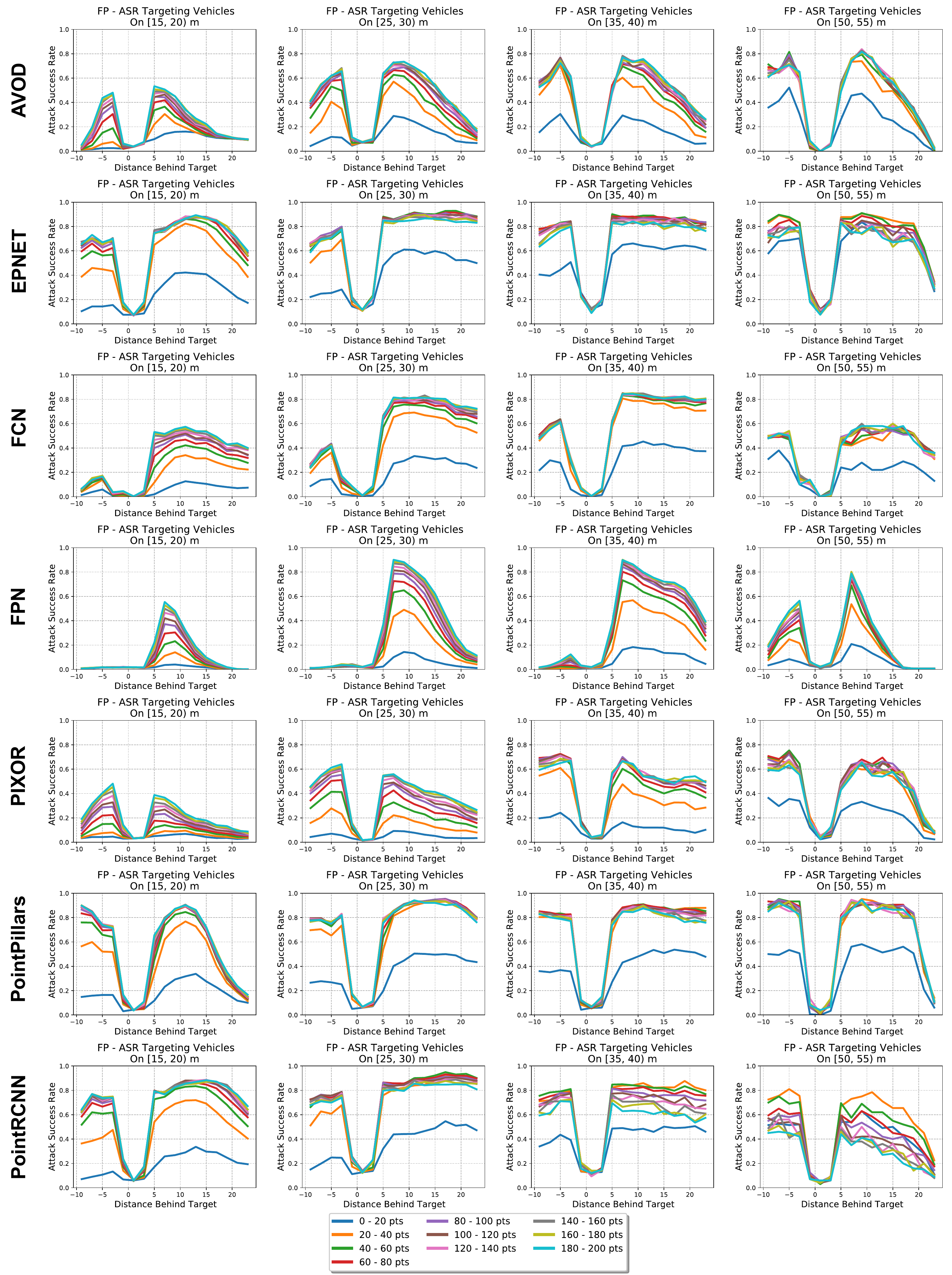}
    \vspace{-8pt}
    \caption{
    Frustum attack success rate against all perception algorithm for different parameters (i.e., number of spoofed points, target vehicle range, distance of spoofing behind the target) combinations, tested on all objects in KITTI validation set. Each tested algorithm (row) is widely vulnerable to the frustum attack. 
    ASR depends on the range to the target vehicle (column). 
    Note a dead-zone near the target vehicle (i.e.~relative distance=0) where attacks do not succeed and increased ASR as target range increases. Most algorithms are vulnerable to spoofing both in front ($<0$ on x-axis) and behind ($>0$ on x-axis) target objects.}
    \vspace{-4pt}
    \label{fig:attack-dist-all-algorithms}
\end{figure*}

Fig.~\ref{fig:attack-dist-avod}, in Section~\ref{sec:6-frustum-attack-results}, summarizes the frustum attack performance against the AVOD perception algorithm for different parameter combinations. 
Here, we provide the results for all other aforementioned algorithms (Fig.~\ref{fig:attack-dist-all-algorithms}).
%
The majority of algorithms are vulnerable to frustum attacks both in front and behind the targeted object. In general, attack success increases as the range to the target object increases (left to right in a row), 
with low attack success for all algorithms when attacks occur near the target object; this is expected as the original object and false positive will be "merged" (i.e.~only a single detection) once they are on top of each other.



\begin{figure}[!t]
    \begin{subfigure}[c]{\linewidth}
      \centering
      \includegraphics[width=0.5\textwidth]{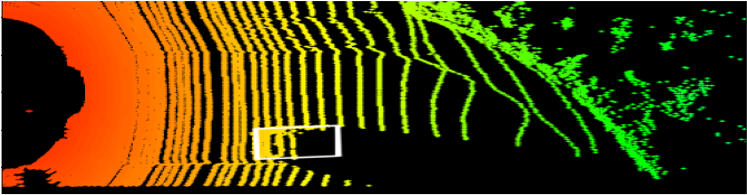}  
      \vspace{-6pt}
      \caption{Full point cloud with ground truth object box at $16~m$ (white)}
    \end{subfigure}
    \begin{subfigure}[c]{\linewidth}
      \centering
      \includegraphics[width=0.5\textwidth]{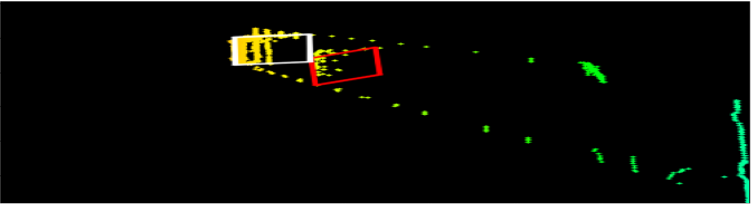}
       \vspace{-6pt}
    \caption{Frustum with original and spoof points - detection (red) at $20~m$}
    \end{subfigure}
    \begin{subfigure}[c]{\linewidth}
      \centering
      \includegraphics[width=0.5\textwidth]{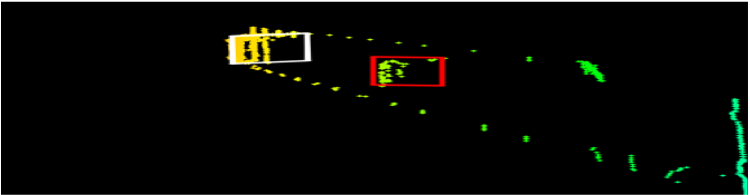}
       \vspace{-6pt}
      \caption{Frustum with original and spoof points - detection (red) at $24~m$}
    \end{subfigure}
    \begin{subfigure}[c]{\linewidth}
      \centering
      \includegraphics[width=0.5\textwidth]{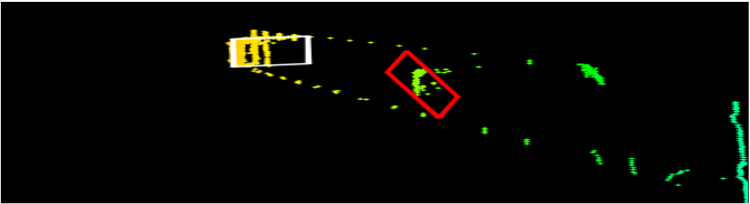}
       \vspace{-6pt}
      \caption{Frustum with original and spoof points - detection (red) at $27~m$}
    \end{subfigure}
 \vspace{-8pt}
  \caption{
  Target object (front, white box) at $16~m$ with 492 points in bounding box. Just 65 points alter the target vehicle’s location and achieve detections (translations) using FPN.}
  \vspace{-6pt}
  \Description{Tracking motivation from BEV.}
  \label{fig:frustum-longitudinal-visualization}
\end{figure}

\vspace{-4pt}
\section{Longitudinal Frustum Attack Visualizations}
\label{sec:appendix-tracking-case-studies}
\vspace{-4pt}

Spoofing points in successively changing distances 
causes the FP injection to appear to travel longitudinally. Fig.~\ref{fig:frustum-longitudinal-visualization} shows a BEV visualization of such a longitudinal attack where the perception 
detects motion of the spoofed points (red).

\end{document}